\documentclass[twocolumn]{aastex63_ionthinspace}

\usepackage{xcolor}
\usepackage{amsmath}
\usepackage{affils}
\usepackage{enumitem}

\newcommand{\HI}{\ion{H}{1}}
\newcommand{\HII}{\ion{H}{2}}
\newcommand{\HH}{${\rm H_2}$}

\newcommand{\OII}{\ion{O}{2}}
\newcommand{\OIII}{\ion{O}{3}}
\newcommand{\halpha}[0]{H$\alpha$}
\newcommand{\hbeta}[0]{H$\beta$}
\newcommand{\te}[0]{$ T_e$}
\newcommand{\nelec}[0]{$n_e$}
\newcommand{\telec}{$T_e$}
\newcommand{\ftst}[0]{[\ion{O}{3}]$\lambda$4363\AA{}}
\newcommand{\foos}[0]{[\ion{O}{3}]$\lambda$5007\AA{}}
\newcommand{\oiidoublet}[0]{[\ion{O}{2}]$\lambda\lambda$3727,3729\AA{}}
\newcommand{\sttoB}[0]{[\ion{O}{2}]$\lambda$7320\AA{}}
\newcommand{\sttoR}[0]{[\ion{O}{2}]$\lambda$7330\AA{}}
\newcommand{\etam}[1][]{\eta_{\rm m #1}}
\newcommand{\asfh}{a_{\rm SFH}}
\newcommand{\fhi}{f_{21}}
\newcommand{\ionline}[3]{[\ion{#1}{#2}]$\lambda$#3\AA{}}
\newcommand{\cc}[0]{cm$^{-3}$}
\newcommand{\lineflux}[1]{F({\rm #1})}
\newcommand{\rrr}[1]{{\textcolor{black}{#1}}}
\newcommand{\srr}[1]{\textcolor{black}{#1}}

\newcommand{\kms}{{\rm km\ s^{-1}}}
\newcommand{\tlbmax}[0]{$\sim$2.5 Gyr}
\newcommand{\tlb}[0]{$t_{lb}$}
\newcommand{\logmstar}[1][]{$\log_{10}(M_\star/M_\odot)$#1}

\newcommand{\lha}{$L_{\rm H\alpha}$}
\newcommand{\note}[1]{\textcolor{purple}{#1}}

\newcommand{\citeprep}[1]{(#1 et al., in preparation)}
\newcommand{\tth}{\textsuperscript{th}}

\defcitealias{otherpaper}{Paper II}

\newcommand{\iffirst}[1]{#1}
\newcommand{\ifsecond}[1]{}

\newcommand{\aat}{AAT/2dF}
\newcommand{\mmt}{MMT/Hectospec}
\newcommand{\sagabg}{SAGAbg-A}

\newcommand{\nemission}{18} 
\newcommand{\ncontinuum}{10} 

\newcommand{\etamest}{$\etam{}=0.92^{+1.76}_{-0.74}$}
\newcommand{\nauroral}{120}
\newcommand{\nfinal}{11925}

\newcommand{\nsagaaatmmt}{11195}
\newcommand{\nsagagama}{180}
\newcommand{\nsagasdss}{550}
\newcommand{\nreference}{698}
\newcommand{\nrefauroral}{23}

\graphicspath{{./}{PaperI_figures/}}

\begin{document}
\shortauthors{Kado-Fong et al.}

\title{SAGAbg I: A Near-Unity Mass Loading Factor in Low-Mass Galaxies via their Low-Redshift Evolution in Stellar Mass, Oxygen Abundance, and
Star Formation Rate}

\author[0000-0002-0332-177X]{Erin Kado-Fong}
\affiliation{Physics Department, Yale Center for Astronomy \& Astrophysics, PO Box 208120, New Haven, CT 06520, USA}

\author[0000-0002-7007-9725]{Marla Geha}
\affiliation{Department of Astronomy, Yale University, New Haven, CT 06520, USA}

\author[0000-0002-1200-0820]{Yao-Yuan Mao}
\affiliation{Department of Physics and Astronomy, University of Utah, Salt Lake City, UT 84112, USA}

\author[0000-0002-4739-046X]{Mithi A. C. de los Reyes}
\affiliation{Department of Physics and Astronomy, Amherst College, 25 East Drive, Amherst, MA 01002}

\author[0000-0003-2229-011X]{Risa H. Wechsler}
\affiliation{Kavli Institute for Particle Astrophysics and Cosmology and Department of Physics, Stanford University, Stanford, CA 94305, USA}
\affiliation{SLAC National Accelerator Laboratory, Menlo Park, CA 94025, USA}

\author[0000-0002-8320-2198]{Yasmeen Asali}
\affiliation{Department of Astronomy, Yale University, New Haven, CT 06520, USA}

\author[0000-0002-3204-1742]{Nitya Kallivayalil}
\affiliation{Department of Astronomy, University of Virginia, 530 McCormick Road, Charlottesville, VA 22904, USA}

\author[0000-0002-1182-3825]{Ethan O. Nadler}
\affiliation{Carnegie Observatories, 813 Santa Barbara Street, Pasadena, CA 91101, USA}
\affiliation{Department of Physics \& Astronomy, University of Southern California, Los Angeles, CA, 90007, USA}

\author[0000-0002-9599-310X]{Erik J. Tollerud}
\affiliation{Space Telescope Science Institute, 3700 San Martin Drive, Baltimore, MD 21218, USA}

\author[0000-0001-6065-7483]{Benjamin Weiner}
\affiliation{Department of Astronomy and Steward Observatory, University of Arizona, Tucson, AZ 85721, USA}
\correspondingauthor{Erin Kado-Fong} 
\email{erin.kado-fong@yale.edu}
  
  \date{\today}

\begin{abstract}
Measuring the relation between star formation and galactic winds is observationally difficult. 
In this work we
make an indirect measurement of the mass loading factor (the ratio between mass outflow rate and star formation rate)
in low-mass galaxies using a differential approach to modeling the 
low-redshift evolution of the star-forming main sequence and mass--metallicity relation. 
We use the SAGA (Satellites Around Galactic Analogs) background galaxies, 
those spectra observed by the SAGA survey that are not associated with the main SAGA host galaxies, to construct a sample of \nfinal{} spectroscopically confirmed low-mass galaxies from $0.01\lesssim z \leq 0.21$ and measure a auroral line metallicity for
\nauroral{} galaxies.
The crux of the method is to 
use the lowest redshift galaxies as the boundary condition of our model, and to infer a mass-loading factor for the sample by comparing the expected evolution of the low \rrr{redshift} reference sample in stellar mass, gas-phase metallicity, and star formation rate against the observed properties of the sample at higher redshift.
We infer a mass-loading factor of \etamest{}, which is in line with 
direct measurements of the mass-loading factor 
from the literature despite the drastically different set of assumptions needed for each approach. 
While our estimate of the mass-loading factor is in good agreement with recent galaxy simulations that focus on resolving the dynamics of the interstellar medium, it is smaller by over an order of 
magnitude than the mass-loading factor produced by many contemporary cosmological simulations. 
\end{abstract}

\section{Introduction}
Low-mass ($M_\star \lesssim 10^{10} M_\odot$) and dwarf ($M_\star \lesssim 10^9 M_\odot$) galaxies 
are systems of interest for topics that range from stellar feedback to large-scale structure to the nature of dark matter.  
In the nearby Universe, ongoing surveys are working to map out the low-mass satellite population of galaxies like our Milky Way in an effort to systematically explore the 
effect of environment on low-mass galaxy evolution and the implication of extragalactic satellite populations on ideas of near-field cosmology \citep{geha2017,mao2021, carlsten2022}. At larger scales, the low-mass end of the stellar-to-halo mass relation 
remains broadly unconstrained but necessary for a full understanding of cosmological structure formation \citep{grossauer2015, leauthaud2020, carlsten2021}, the galaxy-halo connection \citep{nadler2020,danieli2022},
and the properties of dark matter \citep{nadler2021,newton2021}. 
The number densities and density profiles of the dwarfs also
have the potential to shed light on the nature of dark matter \citep[see, e.g.][]{governato2012, delpopolo2016,  bullock2017, nadler2020}. On galactic scales, it has long been thought that 
the structure and evolution of dwarf galaxies are expected to be more sensitive to the nature of 
star formation feedback than are their massive counterparts, making these small systems important observational probes of our understanding of feedback and the baryon cycle \citep[see, e.g.][]{dekel1986, white1991, dalcanton2007, geha2012, onorbe2015, elbadry2016, wetzel2016, hu2019, semenov2021, kadofong2022c, kadofong2022a, jahn2022}.




Star formation and the (self-)regulation thereof are thought be a crucial driver of structure and evolution in low-mass galaxies \citep{kim2018,hu2019,steinwandel2022a,steinwandel2022b}, but quantifying the impacts of the star 
formation cycle on individual galaxies is difficult due to the stochastic and temporally variable nature of star formation. 
\rrr{A crucial component of the self-regulation of star formation is the ability of star formation to remove mass from the gas reservoir of the gas galaxy not only through conversion of gas to stars, but via galactic-scale outflows driven by star formation feedback \cite[see, e.g.][]{lyndenbell1975, pagel1975, carigi1994, lilly2013, muratov2015, anglesalcazar2017, kim2018, hu2019, nelson2019, pandya2020, pandya2021, carr2022, ostriker2022, steinwandel2022a, steinwandel2022b, steinwandel2023_metalloading}.}
The mass-loading factor, $\etam$, is a key metric of star formation feedback that 
quantifies the amount of gas mass lifted out of a galaxy's interstellar medium (ISM) per unit star formation, i.e.:
\begin{equation}
  \etam{} \equiv \frac{\dot M_{\rm out}}{\text{SFR}},
\end{equation}
where $\dot M_{\rm out}$ is the rate of mass outflow from the galaxy. Crucially for this work, 
the mass-loading factor is thought to play a key role in setting the mass--metallicity relation by 
controlling the flow of metal-enriched gas out of the ISM.

To measure a mass outflow rate one must determine both the location of the gas (i.e. that the relevant gas has been lifted out of the galaxy) and the speed of the gas 
(to estimate the rate at which the gas is flowing out of the host galaxy). Due to the projected nature of extragalactic astronomy, it is \rrr{often} difficult to measure both the distance from the midplane $z$ (or $r$) and outflow velocity $v_z$ (or $v_r$). \rrr{Even when both position and velocity can be measured \citep{marasco2023}, the complex dynamical state of the gas reservoirs in low-mass galaxies make the interpretation of their kinematics -- and thus the estimation of the mass-loading factor -- a complex task.} 
Deriving an estimate of a total 
mass density from observable quantities of specific lines is also subject to uncertainties due to assumptions regarding ionization and abundance corrections.
One either observes that the gas has been displaced but not the speed of the displaced gas \citep{mcquinn2019} or observes the velocity of the gas but not its position \citep{martin1999, heckman2015, chisholm2017, marasco2023}.

Another approach to constraining the mass-loading factor is to determine the range of mass-loading factors that are permissible given the current state of the mass metallicity relation or other scaling relations \citep{lilly2013, lin2022}. \rrr{Using the chemical enrichment history of individual or sets of galaxies is a well-established method of modeling their gas cycling processes due to the inherent link between a galaxy's gaseous and stellar content \citep[see, e.g.][]{lyndenbell1975, pagel1975, carigi1994, matteucci2016, maiolino2019, matteucci2021}.}
The difficulty with this more indirect approach is that the mass-loading factor is not the sole parameter that sets the mass--metallicity relation at $z\sim 0$ and some degree of complexity is required to instantiate $z\sim0$ observations; these models have nevertheless been successful in establishing mass-loading factor estimates across a large range of stellar masses \rrr{\citep[see, for example,][]{bouche2010, yin2011, dave2012, lilly2013, yin2023}}.

In this work, we adapt the indirect method of estimating the mass-loading factor 
via its effect on the low-redshift ($z\lesssim 0.2$) evolution of both the star-forming main sequence and mass--metallicity relations. Rather than taking on the substantial hurdle of producing a realistic $z\sim 0$ galaxy population, we will use our observed lowest redshift galaxy sample as our $z=0$ boundary condition.

This approach differs from classic literature methods in its \rrr{intrasample} differential nature\rrr{, where we will explicitly model the selection function of a single sample rather than making a multi-sample comparison \citep[as in, e.g.][]{zahid2012},} and \rrr{by using the lowest redshift galaxies in our sample as a tautologically realistic boundary condition for our method \citep[as opposed to full-fledged regulator models such as][]{lilly2013}}. 
Here, the size and depth of our sample allow us to explicitly characterize the selection function of the SAGA background galaxy sample to enable modeling of chemical evolution on an intrasample basis.
The method allows us to partially circumvent the well-known issue of absolute 
calibrations in gas-phase metallicity estimates \citep{kewley2008, kewley2019}, directly
account for photometric and spectroscopic incompleteness during model comparison, and 
use measured star formation rates (SFRs) as model inputs rather than predicting SFR via star 
formation efficiency arguments.

In order to execute this approach, we need a large spectroscopic sample of low-mass galaxies out 
to a sufficiently large redshift such that some physical evolution of the 
population is expected to be detectable.
The Satellites Around Galactic Analogs (SAGA) background galaxies provide 
such a sample.
Although the main science driver of the SAGA Survey is a census of satellite galaxies around Milky Way-like hosts, the vast majority of the spectra collected are of low-mass galaxies at somewhat higher redshift.
The SAGA sample sits in the wider context of 
efforts that have been put into cataloging the dwarf galaxy population within the Local Volume and nearby Universe \citep{dale2009, lee2009a, lee2009b, berg2012, hunter2012, cook2014}. New and ongoing surveys are now systematically mapping out the low-mass galaxy population at $z>0$ \citep{darraghford2022, luo2023}, which has borne out new possibilities for understanding the broader population of dwarf galaxies and contextualizing our Local Volume on a wider scale.

We will argue that the SAGA galaxies imply an average mass-loading factor of $\etam\sim 1$ at $10^{7.5}\lesssim M_\star/M_\odot \lesssim 10^{9.5}$, in agreement with observational measures of the mass-loading factor of individual galaxies and inconsistent with recent simulations that call for large mass-loading factors at low mass in order to produce realistic $z=0$ dwarfs.

We adopt a flat $\Lambda$CDM cosmology with $\Omega_m=0.3$ ($\Omega_\Lambda=0.7$), and
$H_0=70\ \kms{}\ {\rm Mpc^{-1}}$. We use a \cite{kroupa2001} initial mass function (IMF) unless otherwise specified, and convert literature results that use other IMFs to a 
\cite{kroupa2001} IMF
as stated in the text.

\section{\sagabg{} and the SAGA Background Galaxies}
\iffirst{
The SAGA Survey is a spectroscopic search for satellites down to $M_r\sim-12$ around
MW-like hosts at $z\sim 0.01$ \citep{geha2017, mao2021}. Due to the rarity of the satellite galaxies, the significant majority of spectra collected by SAGA are galaxies that are not associated with the SAGA hosts, in order to reach a highly complete survey. Nevertheless, these non-satellite (background) galaxies tend to be low-redshift and relatively low-mass due to the photometric selection used by SAGA \citep{mao2021}.

These background spectra represent a sample of primarily low-mass galaxies down to a limiting magnitude of 
$m_r\sim21$, a magnitude fainter than the Galaxy And Mass Assembly survey\citep[GAMA,][]{baldry2010} with effective exposure times around twice as long. The SAGA background spectra thus provide a 
fainter and deeper look into the low-redshift, low-mass galaxy population than has been 
possible with previous generation spectroscopic surveys. 

In this work, we consider a subset of SAGA background galaxies which we will call the \sagabg{} sample. 
The nomenclature simply refers to the galaxies' status as background galaxies (bg), and that this is the first 
subset of the background galaxies used for science (-A). The \sagabg{} sample  only includes galaxies for which spectra
were either drawn from archival SDSS or GAMA observations or obtained with \aat{} or \mmt{} 
during the SAGA spectroscopic campaign, and excludes the satellites of the SAGA hosts themselves (i.e., the main science targets of the  SAGA Survey). 
The \sagabg{} sample considered covers an on-sky area of around 110 sq.~deg., though as we will discuss in \autoref{s:modeling} the spatial coverage over this area is not uniform. The data set includes the data presented in~\cite{geha2017} and~\cite{mao2021}, as well as 
additional survey data that will be presented in a forthcoming work \citeprep{Mao}.
}
\ifsecond{
The sample considered in this work is identical to that used in~\citetalias{otherpaper}.
We will reiterate the relevant facets of the sample selection such that the reader may
understand the present work as a standalone, but direct the reader interested in 
further technical detail to~\citetalias{otherpaper}. 
A reader unfamiliar with
\citetalias{otherpaper} will find all the details necessary to understand the work at hand in the proceeding sections.

Although the SAGA Survey is a spectroscopic search for satellites down to $M_V\sim-12$ around
MW-like hosts at $z\sim 0.01$ \citep{geha2017, mao2021}, most of the spectra that have been collected as part of the SAGA Survey are low-redshift, low-mass background galaxies.
In this work, we consider \nsagaaatmmt{} galaxies for which spectra
were obtained with \aat{} or \mmt{} 
during the SAGA spectroscopic campaign 
with an average integration time of 2 hours on the 3.9m AAT and 1.5 hours on the 6.5m MMT. 
}

\begin{figure}[t]
  \centering     
  \includegraphics[width=\linewidth]{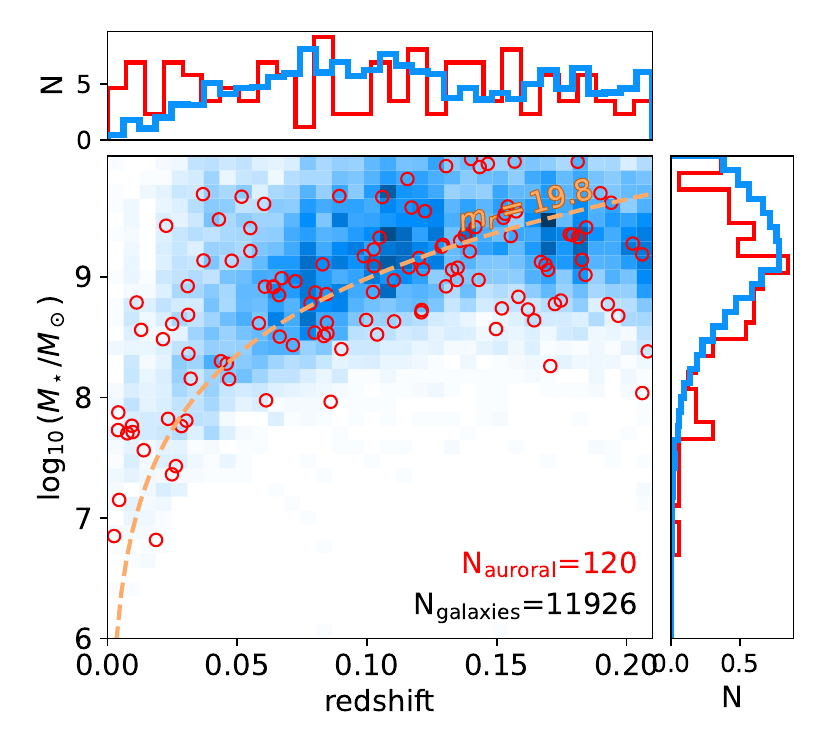}
  \caption{ 
      The distribution of the \sagabg{} sample in redshift and stellar mass; we restrict our sample to those galaxies with stellar masses 
      \logmstar[$<10$] and $z<0.21$. Unfilled red circles show galaxies 
      for which we can measure weak auroral line metallicities, as will be 
      presented in detail in \autoref{s:methods:spectra}.
      The top and right panels show the projection of the distribution over stellar mass and redshift, respectively\rrr{, for both the full (blue) and auroral line-detected (red) samples}.
      In the main panel, the gold dashed curve shows the stellar mass corresponding to an apparent magnitude of $m_r=19.8$ assuming $(g-r)=0.3$, which corresponds
      to the limiting magnitude of the majority of GAMA spectroscopic fields.
      }\label{f:zlogm}
\end{figure}

\begin{figure*}[t]
  \centering     
  \includegraphics[width=\linewidth]{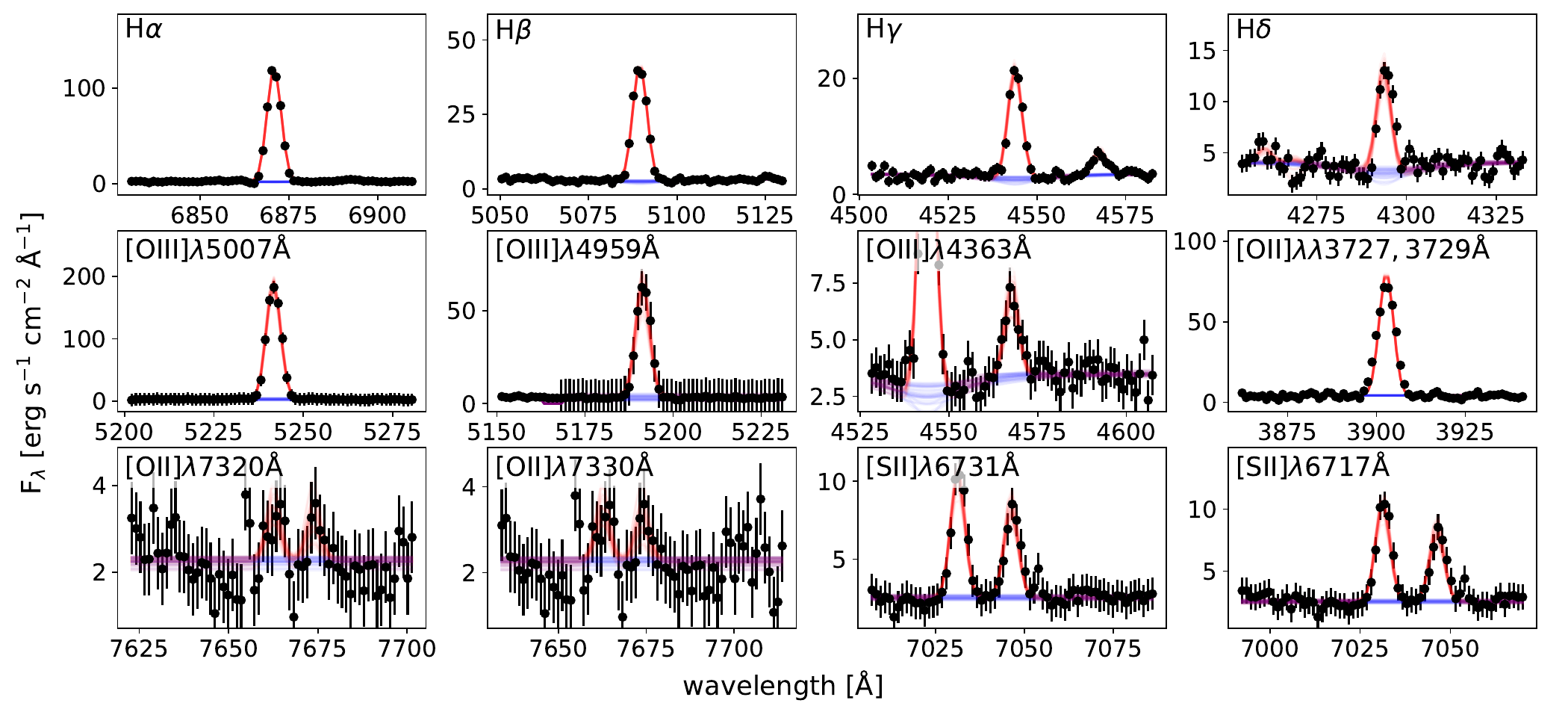}
  \caption{ 
      An example spectrum fit for a galaxy at $z=0.047$ (RA=11h16m05.2s, Dec=+48d20m15.1s), showing fits to different emission lines. Note that the redder \OII{} weak lines are not fit for the galaxy as they are redshifted beyond 8000~\AA{}. In each panel, the top left label indicates the emission line of interest. Black points show 
      observed specific flux, while the red curve shows the best-fit model. The blue curve shows the same best-fit model without emission (i.e., showing only stellar absorption and continuum).      
      }\label{f:linefit_example}
\end{figure*}

\begin{figure*}[t]
  \centering     
  \includegraphics[width=\linewidth]{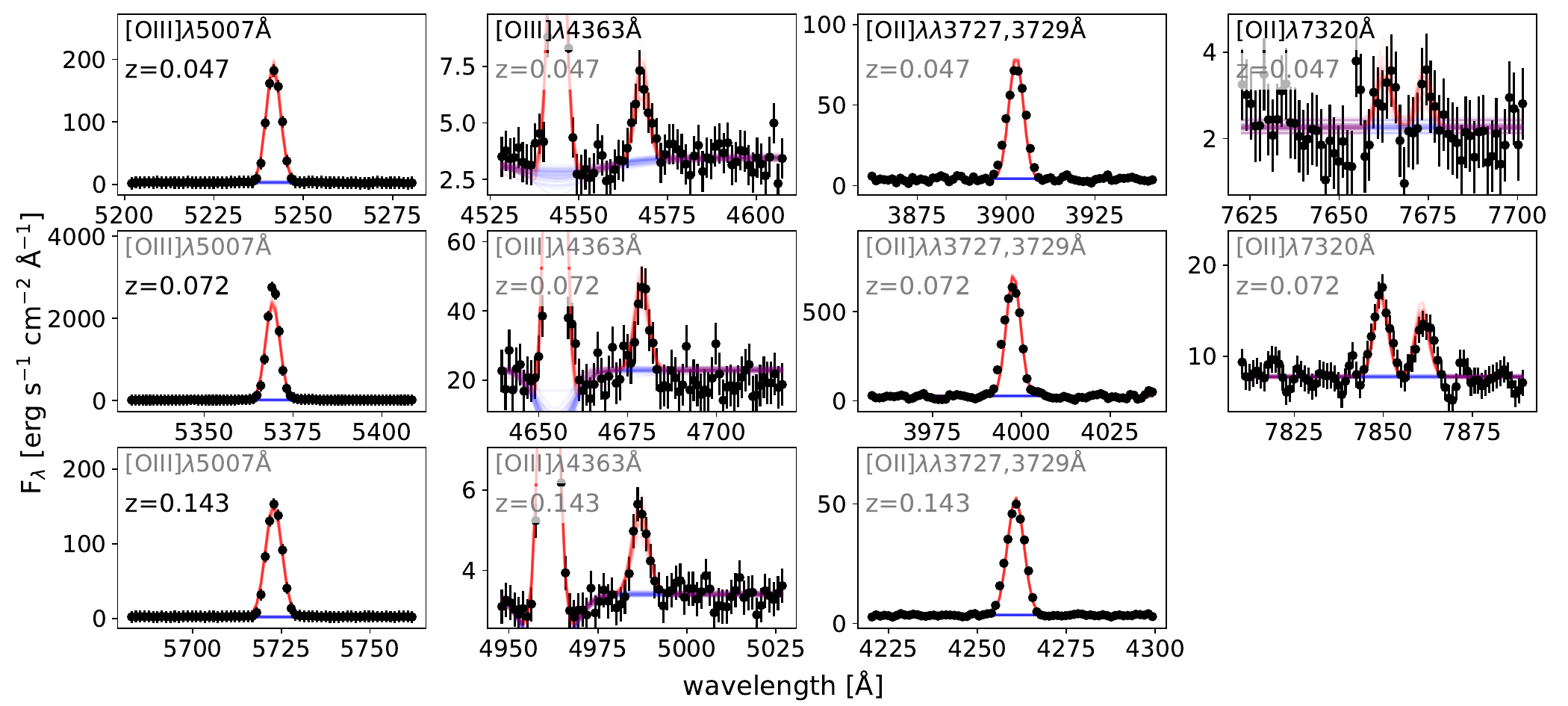}
  \caption{ 
      The same as \autoref{f:linefit_example}, but showing only 
      the key weak auroral lines (\ionline{O}{3}{4363} and \ionline{O}{2}{7320},\ionline{O}{2}{7330}) and their strong line counterparts for three galaxies at different redshifts. Note that in the highest redshift example shown here (bottom row), the \ion{O}{2} auroral lines are redshifted out of the spectrum wavelength range and are hence not shown.
      The galaxies shown are at (RA=11h16m05.2s, Dec=+48d20m15.1s; RA=21h39m13.5s, Dec=-42d35m07.3s, RA=13h46m56.4s, Dec=+41d54m26.0s).
      }\label{f:keylines_example}
\end{figure*}

\iffirst{
  \subsection{Photometric Selection}\label{s:sample:photometric}
The photometric selection of the SAGA Survey is designed to completely span the 
range of photometric properties occupied by the SAGA satellites, which are low-mass (\logmstar[$\lesssim 10$]) galaxies at $z\lesssim 0.013$. As the SAGA Survey progressed, the photometric selection has evolved to exclude more background galaxies while maintaining the completeness of satellite galaxies \citep{geha2017, mao2021}. 
As such, when we consider the SAGA background galaxies, their effective photometric selection is not uniform across different SAGA hosts. 
Because we are interested \textit{only} in the galaxies not associated with the SAGA hosts, however, we argue that it is sufficient to characterize the realized aggregate selection function of the SAGA background galaxies. 

{}

The SAGA host fields, when the satellites themselves are excluded, can be considered to be random fields. 
Since there should be no differences in the population of galaxies between two random fields, there should also be no difference -- modulo cosmic variation -- between targeting each random field with somewhat different photometric cuts and targeting \rrr{all} fields with the average photometric cuts as weighted by number of spectra obtained using each selection function.

Taking this aggregate approach allows us to characterize the \sagabg{} sample by its effective
photometric selection, which we will quantify as the fraction of SAGA photometric targets that 
are in the \sagabg{} sample (we remind the reader that this is not equivalent to the fraction of galaxies with a SAGA redshift). 
Our model is built around the idea that our sample is incomplete, and that we can quantify our incompleteness relative to some ``most complete'' subset. We will return to this idea more quantitatively in \autoref{s:modeling} when we build a model to understand the physical evolution of our observed samples.

While we defer a 
full characterization of this aggregate selection function and its impact on the present sample to 
\autoref{s:modeling}, we would like to acquaint the reader with the sample at hand we 
show in \autoref{f:zlogm} the distribution of the sample over redshift and
 stellar mass. The stellar mass
that would correspond to an apparent magnitude of $m_r=19.8$ for a galaxy with 
a restframe color of $(g-r)=0.3$ using our color-mass relation as an orange curve. 
This apparent magnitude is roughly equal to the limiting magnitude of most field in the
Galaxy Mass and Assembly (GAMA) survey \citep{driver2009}.


}
\ifsecond{
  \input{./targeting_second.tex}
}
\iffirst{
      \subsection{SAGA Spectra}\label{s:methods:spectra}
    Measuring star formation rates from \halpha{} luminosities and gas-phase metallicites from temperature-sensitive oxygen line flux ratios requires flux-calibrated spectra.
    As mentioned earlier, the SAGAbg-A sample used in this work includes all spectra obtained from \mmt{} and \aat{} (including GAMA archival
    spectra), as well as all archival spectra of SAGA photometric targets taken by SDSS.
    
    We find that the relative flux calibration of the SAGA AAT and MMT spectra, particularly those spectra from \aat{}, degrades at $\lambda \gtrsim 8000$ \AA{}. \rrr{This cutoff was determined by comparing SAGA spectra to spectra of the same galaxies released by the GAMA survey; we deem the SAGA flux calibration unreliable when the fractional difference $f(\lambda) \equiv |F_\lambda^{\rm SAGA}(\lambda) - F_{\lambda}^{\rm GAMA}(\lambda)|/\sqrt{\sigma_{F_\lambda}^{\rm SAGA}(\lambda)^2+\sigma_{F_\lambda}^{\rm GAMA}(\lambda)^2}$ between the SAGA and GAMA flux calibration reaches $f(\lambda_{\rm cut})>2$ and the sign of $df/d\lambda$ is constant at $\lambda>\lambda_{\rm cut}$. }
    We thus cut off the spectra in our analysis at $\lambda\rrr{_{\rm cut}}=8000$ \AA{} for spectra originating from \aat{} and $\lambda\rrr{_{\rm cut}}=8200$ \AA{} for \mmt{}. 
    We note that the spectra obtained by SAGA have an average integration time of 2 hours on the 3.9m AAT and 1.5 hours on the 6.5m MMT and comprise the vast majority (94\%) of the final sample. 
    
    Next, we flux calibrate our spectra using the SAGA $g-$ and $r-$band photometry by assuming that the spectrum is representative of the full galaxy (i.e. assuming that there are no underlying population gradients) and applying the multiplicative conversion to reproduce the SAGA photometry from integrating the SAGA 
    spectrum. This method is well-suited for our sample because the galaxies are generally small even compared to the relatively small \aat{} and \mmt{} fibers: the median on-sky effective radius of our sample is 1.2\arcsec{}, while the \aat{} and \mmt{} fibers are 2.1\arcsec{} and 1.5\arcsec{} in diameter, respectively. 
    Because the $g-$ and $r-$bands lie on different \aat{} spectrograph arms, we allow for different flux calibrations in $g$ and $r$ for \aat{} spectra to correct for any discontinuities between the two arms 
    that persisted beyond relative flux calibration. Finally, because the relative flux calibration is quite important for our metallicity measurements, we remove any \aat{} spectra that contain a discontinuity of greater than 5\% across the spectrograph break after absolute flux calibration. 
    
    We apply flux calibration to the SDSS spectra in the same way that we do for the SAGA AAT and 
    MMT spectra. The flux calibration provided by the GAMA public data release are already 
    calibrated consistently to our flux calibration and so we do not perform an additional 
    flux calibration. For all spectra we correct for galactic extinction assuming a 
    \cite{cardelli1989} extinction curve with an assumed $R_V=3.1$ and the galactic $A_V$ measured by
    \cite{schlafly2011} at the position of each target via the IRSA galactic dust and reddening server \citep{irsadust}. 
    
}
\ifsecond{
  \input{./spectra_second.tex}
}
\section{Measurements}\label{s:measurements}
\iffirst{
In this work, we will infer weak line ratios and strong line fluxes of the 
\sagabg{} spectra, convert those line measurements to estimates of ISM
physical conditions, and use those physical conditions to make an inference about
the evolution of the galaxy sample over a short period of cosmic time (e.g. lookback times of less than \tlb{}$\lesssim 2.5$ Gyr).
Throughout this process, we will leverage our external knowledge (e.g. of atomic physics, of the
sample selection) to make an inference of the system as a whole rather than 
independently estimating parameters from individual components of our data. 
}

We will measure gas-phase metallicities for the sample using the ``\rrr{electron} temperature''
approach, which uses temperature-sensitive weak line ratios to constrain the temperature
(and thus the emissivity) of \OII{} and \OIII{}-emitting gas.
The quality of the SAGA spectra allows us to confidently measure these key auroral line fluxes
(\ionline{O}{3}{4363}, \ionline{O}{2}{7320}, and \ionline{O}{2}{7330}) for a small but significant
fraction of the galaxy sample. These line ratios 
are strongly temperature-dependent and density-insensitive (for the 
densities relevant to this work). Auroral line metallicities are often thought of as ``gold standard'' metallicities because ion abundance can be 
directly computed from flux ratios (e.g. of \ionline{O}{3}{5007}/H$\beta$) once the temperature and
density (and thus the line emissivities) are measured from temperature-sensitive line ratios. 
\rrr{However, i}t is important to note that auroral line metallicities also suffer from biases. The most often-used auroral line \ionline{O}{3}{4363} is rarely seen at 
metallicities exceeding $\log_{10}\text{(O/H)}+12 = 8.7$ --- this bias is mitigated by the low metallicity of our sample 
\citep[for a recent review, see][]{kewley2019}. 
\rrr{It is also known that auroral line metallicity samples tend to preferentially select metal-poor galaxies \citep{kewley2008}, and that samples selected to have \ionline{O}{3}{4363} detections can result in an underestimate of the mass-metallicity relation \citep{curti2020}. Furthermore, the impact of temperature and density inhomogeneities on \telec{}-based metallicity estimates remains an open question at both the \ion{H}{2} region and galactic scale \citep{chen2023_mrk71, mendezdelgado2023, mendezdelgado2023_mrk71}.}

Nevertheless, \rrr{we find that it is}
preferable to compare auroral line metallicities across cosmic time than to compare either empirical calibrations of strong line ratios or theoretical metallicity calibrations. The former regularly show mean offsets of 0.5--1.0 dex between theoretical calibrations of the same strong line ratio, and the latter's fidelity relies on our uncertain understanding of stellar atmospheres, evolution, and other model assumptions. \rrr{In \autoref{s:appendix:strongline}, we rerun our analysis using various strong line metallicity calibrations from the literature and find that the conclusions are robust against choices of metallicity calibrator that accurately reproduces the known mass-metallicity relation of low-mass galaxies in the nearby Universe.}

\subsection{Line Measurements}
\iffirst{
  The first step to measuring gas-phase metallicities is to measure fluxes of the relevant lines in our
sample. The most important lines for our analysis in this paper series are \halpha, \hbeta, 
\ftst, \foos{}, \oiidoublet, \sttoB, and \sttoR. We also fit a wider set of lines 
including the density-sensitive pair \ionline{S}{2}{6717} and \ionline{S}{2}{6731}. We show an example of a SAGA spectrum around the emission lines of interest in \autoref{f:linefit_example} and several examples of the key weak auroral lines in galaxies across the redshift range of our sample in \autoref{f:keylines_example}. A full accounting of the lines fit for this analysis will be provided as a value-added catalog in a
forthcoming work \citeprep{Mao}.

We fit our emission lines and Balmer absorption features simultaneously; we
allow the amplitude and position to vary for each line but assume that all the emission lines can be well-described 
by a Gaussian profile of the same width $\sigma_{em}$. The Balmer absorption features are also fit as an ensemble of 
Gaussian profiles wherein we require that the ratio between the equivalent width of each absorption feature and the
equivalent width of the H$\beta$ absorption feature is held to be unity for H$\gamma$ and H$\delta$, and 
0.5 for H$\alpha$ (i.e. $\rm EW_{H\alpha} = 0.5 EW_{H\beta}$). This relationship is chosen based on the models of 
\cite{gonzalezdelgado1999}; we adopt a strict assumption here because the H$\alpha$ absorption is not spectrally 
resolved and is therefore degenerate with the emission component. We approximate the
continuum local to each emission line as a constant value, and hold the continuum flux of lines separated by less than
140\AA{} to be equal. 
We infer the properties of the all the emission line profiles in each 
SAGA spectra simultaneously via the Markov Chain Monte Carlo ensemble sampler implemented in \textsf{emcee}
\citep{emcee}.

Our approach allows us to straightforwardly incorporate our knowledge of atomic physics into 
the line profile fitting by including this information into our formulation of the prior. In particular, we have strong 
constraints on the minimum allowable flux ratio of lines originating from the same species 
wherein the redder line is in the numerator. As an example, consider the flux ratio between \halpha{} and \hbeta{}, 
i.e., the Balmer decrement.
For a gas at \nelec{}$=10^3$ \cc{} and \te{}$=10^4$ K, the intrinsic flux ratio between \halpha{} and \hbeta{} should 
be 2.86. In the presence of reddening from dust, the observed flux ratio may be larger than this value (but not smaller). 
We thus adopt a sigmoid prior
\begin{equation}
  \rm Pr(A_{H\alpha}, A_{H\beta}) = (1 + e^{-k(A_{H\alpha}/A_{H\beta} - \tilde A_{H\alpha,H\beta}b)})^{-1}
\end{equation} 
where
$k=30$ and $b=0.925$ are the shape parameters of the sigmoid function, $\tilde A_{H\alpha,H\beta}$ is the intrinsic
flux ratio (here assumed to be 2.86), and $\rm A_{H\alpha}$ and $\rm A_{H\beta}$ are the
amplitudes of the \halpha{} and \hbeta{} lines, taking advantage of the fact that the ratio of the fluxes will be 
equal to the ratio of the amplitudes when the linewidth is equal. Note that the prior allows for, but is weighted against, 
line ratios somewhat below the assumed minimum intrinsic ratio.
The same exercise can be performed with the
other lines in the spectrum; we assume sigmoid priors with the following assumed intrinsic ratios:
\begin{itemize}[itemsep=-0.5mm]\vspace{-2mm}
  \item[] $\tilde A_{H\alpha,H\gamma}=6.11$ 
  \item[] $\tilde A_{H\alpha,H\delta}=11.06$ 
  \item[] $\tilde A_{[OIII]5007,[OIII]4959}=2.98$ 
  \item[] $\tilde A_{[OIII]5007,[OIII]4363}=6.25$ 
  \item[] $\tilde A_{[SII]4077,[SII]4069}=6.25$,  
\end{itemize}
based on the minimum intrinsic line ratio for a reasonable ($10^2\lesssim T\lesssim 10^5$ K) range in temperature as 
computed by the package \textsf{pyneb}. 

We adopt a simple Gaussian likelihood over our model parameters:
\begin{equation}
  \ln\mathcal{L}(\vec F_\lambda | \vec \theta_\ell) = -\frac{1}{2}\sum_{i}{\left[\frac{{\left( \hat F_\lambda(\lambda_i,\vec \theta_\ell) - F_{\lambda,i} \right)}^2}{\sigma_{F_{\lambda,i}}^2} +
  \ln{\left( 2 \pi \sigma_{F_\lambda,i}^2\right)}\right]},
\end{equation}   
where the index refers to the index of the resolution element, $\hat F_\lambda(\lambda_i)$ and $F_{\lambda,i}$ are the predicted and observed specific fluxes in that resolution element, respectively, and $\sigma_{F_\lambda}$ is the uncertainty in the specific flux. 
The observed spectrum is thus simply $\vec F_\lambda = \{F_{\lambda,0}, F_{\lambda,1} ... F_{\lambda, M}\}$ for $M$ resolution elements.
The vector $\vec \theta_\ell$ is of length $2N_{\rm em} + N_{\rm cont} + 3$; $\vec \theta_\ell$ describes the parameters of the line model. Here $N_{\rm em}\leq\nemission{}$ is the number of emission lines fit and $N_{\rm cont}\leq \ncontinuum$
is the number of continuum regions fits. The numbers here are given as upper limits 
because lines that are redshifted beyond $\lambda=8000\text{ \AA}$ are excluded from 
the model.
There are then an additional three parameters: $\sigma_{\rm em}$, the width of the emission lines, $\text{EW}_{\rm abs}$, the 
equivalent width of the Balmer lines, and $\sigma_{\rm abs}$, the width of the (Balmer)
absorption lines. \rrr{We show an example of our line-fitting procedure for a single galaxy in \autoref{f:linefit_example}, and several examples of the key \OII{} and \OIII{} lines we use to estimate gas-phase metallicities across the sample redshift range in \autoref{f:keylines_example}.}

Because some of the lines that we fit are quite weak (at maximum $\sim 10\%$ the flux of \halpha{} and
\ionline{O}{3}{5007}), we visually inspect each galaxy in which the inference predicts a less than 16\% chance that the auroral oxygen lines could originate from stochastic fluctuations to ensure that the measured features are not observational artifacts such as residuals from sky subtraction.
Finally, in \autoref{s:appendix:fluxcalibration} we test both our flux calibration and line measurement methods against published GAMA survey 
results to verify that our methodology is consistent with established literature methods.
}
\ifsecond{
  \input{./lines_second.tex}
}

\begin{figure}[htb]
  \centering     
  \includegraphics[width=\linewidth]{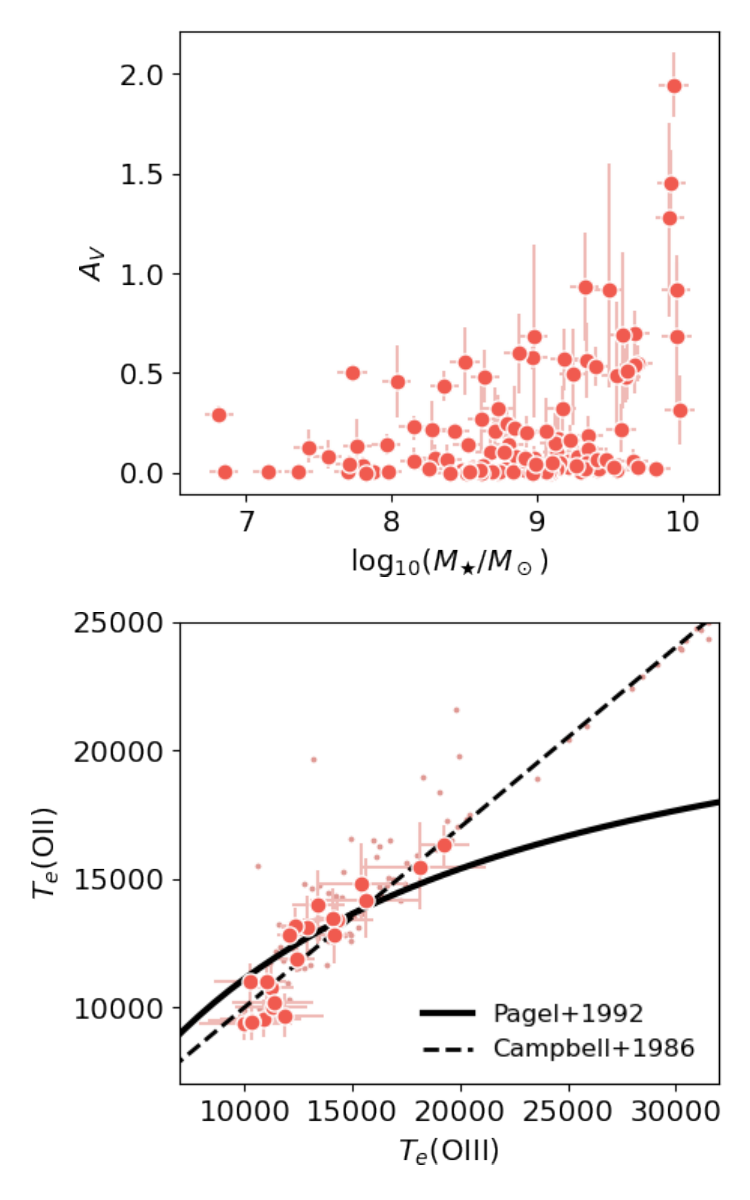}
  \caption{ 
      \textit{Top:} stellar mass versus \rrr{internal} visual extinction ($A_V$) for the sample of galaxies for which we could measure a reliable auroral line metallicity. 
      \textit{Bottom:} \telec(\OIII) versus \telec(\OII), where large points with error bars show galaxies where the auroral 
      lines \ionline{O}{2}{7320} and \ionline{O}{2}{7330} are covered by the portion of the spectra with a reliable flux calibration and
      a strong detection (probability of being drawn from a blank spectrum $<0.05$) of at least one of the red auroral \OII{} lines is present.
      Small scatter points indicate galaxies for which a weak auroral line metallicity could be measured from \ionline{O}{3}{4363}, 
      but for which the \OII{} lines were outside the rest-frame wavelength coverage of the SAGA spectrum. These galaxies are expected to 
      follow the assumed prior temperature relation much more closely, though upper limits on \ionline{O}{2}{7320} and \ionline{O}{2}{7330}
      may still provide information that causes the estimate of \telec(\OII) to deviate from the mean value assumed by the prior relation of \citet[][dashed line]{campbell1986}.
      }\label{f:physicalproperties}
\end{figure}

\subsection{Determining ISM conditions}\label{s:methods:ismconditions}
Line emissivity coefficients are generally a function of temperature (\te{}) and density (\nelec{}); thus, to invert
line flux ratios to abundance ratios, we need to know the physical condition of the gas that
is emitting the lines. Similar to our approach to measuring line fluxes, we will infer our
ISM conditions of interest (\te(\OIII), \te(\OII), and $A_V$) simultaneously via Markov Chain
Monte Carlo sampling implemented from \textsf{emcee}. We additionally adopt a
\cite{calzetti2000} extinction curve and note that the best-fit extinction in these galaxies tends to be low, with $A_V<0.5$, which is unsurprising given their low stellar masses (see \autoref{f:physicalproperties}).

A common method to estimate 
\te{} is to use the auroral  \ftst{} line to leverage
the temperature-sensitivity and density-insensitivity of the \ftst{}/\foos{} flux ratio. This 
allows us to estimate the average \te{} probed by O$^{++}$, which we will follow the
literature in calling the
``high ionization'' (O$^{++}$) zone electron temperature. However, \HII{} regions are not 
typically isothermal --- the average temperature that O$^{++}$ lines probe is not necessarily the 
same as the average temperature that O$^{+}$ lines probe. To constrain the overall oxygen 
abundance, then, one needs a constraint on both \te(\OIII{}) and \te(\OII).
A relation between the high-ionization and ``low-ionization'' (O$^{+}$) zones
is often adopted \citep{berg2012}. 
However, these relations are based on photoionization models of 
\HII{} regions \citep{stasinska1990}, and different models give different relations between
the effective \te{} of each species \citep[see, e.g.][]{campbell1986,stasinska1990,pagel1992}.

Fortunately, there are also temperature-sensitive
\ion{O}{2} line ratios that can provide a 
more direct constraint on \te(\OII), \rrr{though we note that these line ratios do have a moderate dependence on density \citep{hagele2006,perezmontero2017}}. These are \sttoB{}/\oiidoublet{} and \sttoR{}/\oiidoublet{}.
Unfortunately, these lines are close to our red wavelength cutoff (8000\AA{}) even at
$z=0$ and pass out of our redshift range at $z=0.09$. Additionally, at low SNR
it is not guaranteed that both \ftst{} and the pair of weak \OII{} lines will yield
high-confidence detections. To account for either missing coverage or marginal detections, instead of 
measuring \te(\OIII) and \te(\OII) independently we adopt a prior over their relationship based on the model prediction of~\cite{stasinska1990}:
\begin{equation}\label{e:telec}
  \text{\te(\OII)} \sim \mathcal{N}(0.7\ \text{\te(\OIII)} + \text{T}_0, \sigma_{\text{\te}}^2),
\end{equation}
where $\text{T}_0=3000\text{ K}$ and we assume $\sigma_{\text{\te}}=1500\rm\ K$.
Here, $\mathcal{N}(\mu,\sigma^2)$ denotes a normal 
distribution with mean $\mu$ and variance $\sigma^2$. We have also \rrr{performed} the same inference using the
temperature relation of~\cite{pagel1992}\footnote{\cite{pagel1992} propose that ${\text{\te(\OII)}={2\times10^4\ {\rm K}}({(\text{\te(\OIII)}/10^4\rm\ K)^{-1} + 0.8})^{-1}}$, which results in lower \te(\OII) for 
high \te{\OIII} as seen in \autoref{f:physicalproperties}. } and found that while our assumption for the form of 
the prior does result in different estimates of \te(\OII) when \ionline{O}{2}{7220} and \ionline{O}{2}{7330}
are not covered, it does not significantly change our final metallicity estimates given the 
uncertainty on the posterior distribution of gas-phase metallicities.

We adopt uniform priors over \te({\OIII}) and $A_V$:
\begin{equation}
  \begin{split}
    \text{\te(\OIII)}&\sim \mathcal{U}(5\times10^3, 4\times10^4)\ {\rm K}\\
    A_V&\sim \mathcal{U}(0,\srr{100}),
  \end{split}
\end{equation}
where the bounds on the former are set by the range of \te(\OIII) seen in high-SNR measurements 
of individual \HII{} regions in low-mass galaxies by \cite{berg2012}. \rrr{}

We would also like to constrain the density associated with the line emission, \nelec{}. 
However, the density-sensitive line ratio that we have access to is from \ion{S}{2} 
(\ionline{S}{2}{6717}/\ionline{S}{2}{6731}), and is only 
sensitive down to \nelec{}$\sim 10^2$ \cc{}. We have run a version of this inference with 
\nelec{} allowed to vary and the
\ion{S}{2} lines included, and find in practice that our sample is consistent with
the 100 \cc{} low-density limit. 
This is in line with previous works over similar mass ranges
\citep{berg2012, andrews2013}, so we choose to adopt \nelec{}$= 10^2$ \cc{} for this work to avoid
propagating the uncertainty of the \ion{S}{2} density estimate without adding additional 
information to the inference. \rrr{We also find that the inclusion of density as a free parameter does not significantly affect our
inference of \te(\OII) despite the density dependence of the indicator line ratio. This is likely both because the \OII{} line ratio is more sensitive to temperature than density over the densities probed, and because the overall oxygen abundance is expected to be dominated by doubly-ionized oxygen at these metallicities \citep{curti2017}.}

Here we use the posteriors of the line-fitting inference to construct 
a likelihood via density estimation. In particular, we estimate the 
probability density function of our line ratios of interest from the posterior distributions over line fluxes obtained in the previous section via 
a Gaussian kernel density estimate. We perform the density estimation independently for each of the line ratios considered; the 
likelihood is then simply the product of the estimated likelihood for each 
line ratio. \rrr{For each line ratio considered, we first compute the line ratio predicted from the corresponding \telec{} using the flux ratios computed using \textsf{pyneb} using the transition probabilities and temperature-dependent collision strengths reported by \cite{storey2000_oiiiatom} and \cite{storey2014_oiiicoll} for \ion{O}{3}, \cite{zeippen1982_oiiatom} and \cite{kisielius2009_oiicoll} for \ion{O}{2}, and \cite{pequignot1991_hrec} and \cite{storey1995_hrec} for H recombination. We then apply differential reddening assuming a \cite{calzetti2000} extinction curve with an assumed $R_V=4.05$ to determine the expected observed line ratio for a given value of $A_V$.}

In \autoref{f:physicalproperties} we show the distribution of our inferred physical properties in the stellar mass-extinction plane (top) and 
\telec{}(\OIII)-\telec{}(\OII) plane (bottom). Our inferred extinctions show generally low $A_V$ at low mass and increasing maximum observed $A_V$ with stellar mass, in good agreement with previous works in the same mass range \citep[e.g.][]{lee2009b}. Our \telec{}(\OII{}) inferences generally follow the prior defined by \autoref{e:telec}. Even when adopting the \cite{pagel1992} temperature relation as a prior in \autoref{e:telec}, there is some suggestion that the sample is better fit by the \cite{stasinska1990} model (dashed line) at \telec{}(\OIII)$\sim 10^4$ K, though we caution that the difference between the 
\cite{pagel1992} and \cite{stasinska1990} temperature relations are small in this regime.

\subsection{Gas-phase Metallicities}
We follow conventional methods presented in the literature
to derive 
oxygen abundances from our inferred \telec{}(\OIII) and \telec{}(\OII) \citep[see, e.g.][]{berg2012, andrews2013, berg2019} . 
As discussed above, we fix \nelec$=10^2$ \cc{} in our abundance estimates. We compute the
abundances of O$^+$ and O$^{++}$ with respect to H$^{+}$ 
by comparing each species' strong lines with 
respect to \hbeta{} after correcting for extinction contributed by the target galaxy's ISM using a~\cite{calzetti2000} extinction curve and our
inferred $A_V$. 

We use the package \textsf{pyneb} to compute the emissivity of 
each line with respect to \hbeta{}, where:
\begin{equation}
  \frac{N(\rm O^X)}{N(\rm H^+)} = \frac{F^{\rm e}_{{\rm [O^X]}_i}}{F^{\rm e}_{\rm H\beta}} \frac{j_{\rm H\beta}(T_e({\rm H^+}),n_e)}{j_{{\rm [O^X]}_i}(T_e({\rm O^X}),n_e)},
\end{equation}
where $N(.)$ is the column density of the species, and X refers to the ionization state of oxygen. The superscript \textit{e} refers to emitted (i.e., galactic extinction and reddening-corrected) line flux. 
Here we assume that \telec(H\textsuperscript{+}) is equal to the geometric mean of \telec{}(\ion{O}{2}) and \telec(\OIII). For
O\textsuperscript{++} we take the mean of the relative abundances as 
determined independently using \ionline{O}{3}{4959} and \ionline{O}{3}{5007}.
We use the blended \ion{O}{2} doublet to determine the relative column density O\textsuperscript{+} by considering the expected emissivity from both lines in the doublet.

Finally, we follow the typical assumption that the total oxygen abundance 
is the sum of the dominant ionic species abundances relative to H$^{+}$, i.e.:
\begin{equation}
  \rm \frac{O}{H} \approx \frac{O^+}{H^+} + \frac{O^{++}}{H^+},
\end{equation}
which has generally been found to hold true in studies that seek to directly 
detect higher ionization states of oxygen via \ion{O}{4} lines in dwarf galaxies \citep{berg2019}.

\iffirst{
  
\subsection{Star Formation Rates}
We estimate star formation rates following 
\cite{calzetti2013} for a Kroupa IMF:
\begin{equation}\label{e:lha2sfr}
  \text{SFR} = \beta_c {F^{\rm e}_{\text{\halpha{}}} (4\pi d_L^2)},
\end{equation}
where $\beta_c=5.5\times10^{-42}\ M_\odot\text{ erg}^{-1}$ is a scaling factor that 
depends on the adopted IMF and evolutionary library, as well as GMC-scale 
assumptions about ionizing photon leakage from \HII{} regions and ionizing photon 
absorption by dust. We correct for optical extinction from a fit to the Balmer decrement; we 
show the distribution of $A_V$ for the same as a function of stellar mass in \autoref{f:physicalproperties}.
{}

\subsection{Stellar Masses}
We follow the stellar mass prescription established for the SAGA survey 
by \cite{mao2021} and estimate our stellar masses as:
\begin{equation}
  \log_{10}\left( \frac{M_\star}{M_\odot} \right) = 1.254 + 1.098(g-r)_{\rm rf} - 0.4 M_{r,\rm rf}.
\end{equation}
The ``rf'' subscript above indicates restframe measurements. 

This relation is an adaptation of the~\cite{bell2003} mass-to-light and color relation that has been calibrated for the
SAGA sample via a comparison to stellar mass estimates in the SAGA mass range.
As in~\cite{mao2021}, we assume an uncertainty of 0.2 dex for our stellar mass estimates. 

Although the color--stellar mass relation that we use in this work has been calibrated for our stellar mass range by \cite{mao2021}, it is worth noting that color--mass relations for low mass galaxies are often extrapolations from relations calibrated at higher mass. To make sure that the SAGA color--mass relation is applicable to our 
background galaxies, we compare the publicly available stellar masses produced 
by the GAMA survey via SED fitting \citep[\textsf{StellarMassesLambdarv20},][]{taylor2011} with the stellar masses that we would assign 
to these galaxies using the above color--mass relation. Here we use the same photometry 
as the GAMA SED-fit stellar masses to exclude any additional systematic differences 
that could result from a comparison of different photometric methods. 

We find that, over 6.8$<$\logmstar[$\leq$10], there is an average offset of 
$\log_{10}(M_\star/[M_\odot])_{SED} -\log_{10}(M_\star/[M_\odot])_{\rm SAGA} = \Delta$\logmstar[$=0.01$] between the two methods. However, there is a significant 
positive slope to the relation between $\Delta$\logmstar[] and \logmstar[] such that 
lower stellar masses show a larger discrepancy, up to a median 
$\Delta$\logmstar[$=0.13$]
at \logmstar[$<7.2$]. This slope is significant and 
is in line with ongoing work to inspect the fidelity of 
photometric stellar mass estimates at low stellar masses \citeprep{de los Reyes}; for 
the scope of this analysis, however, we will simply emphasize that this offset is 
below our assumed uncertainty in stellar mass of $0.2$ dex.

}
\ifsecond{
  \input{./SFR_Mstar_second.tex}
}

\subsection{Final Sample Numbers}\label{f:measurements:samplesize}
There are 24074 galaxies in the SAGA targeting catalog that lie in our \sagabg{} selection criteria -- i.e. a spectrum from AAT, MMT, GAMA or SDSS at $z<0.21$ with a stellar mass estimate of \logmstar[$<10$]. We discard galaxies with insufficiently accurate relative flux calibrations (see \autoref{s:methods:spectra}), where the uncertainty on the optical extinction was greater than 1 dex, or where the emission lines of interest were contaminated by observational artifacts. Requiring that \hbeta{} is sufficiently well detected to derive a reliable measure of $A_V$ removes a significant number of galaxies for which only the strongest lines (typically \halpha, \oiidoublet{}, and \ionline{O}{3}{5007}) are detected at high SNR.

We thus arrive at a final sample of \nfinal{} low-mass, low-redshift galaxies; of these, we measure a reliable auroral metallicity estimate in \nauroral{}. \rrr{Of these 120 galaxies, 66 have spectra which cover both the \OIII{} and \OII{} auroral lines.} 
The vast majority of our galaxies are from SAGA AAT/MMT observations, which account for \nsagaaatmmt{} spectra. Archival GAMA and SDSS observations account for \nsagagama{} and \nsagasdss{} galaxies, respectively.

\rrr{We include the measured line fluxes, inferred metallicities, and estimated ISM conditions ($A_V$, \te(\OIII), and \te(\OII)) of the \nauroral{} galaxies with reliable auroral line metallicity estimates as a catalog associated with this work. In \autoref{t:catalog} we show an excerpt of the provided catalog.}

{}

\begin{splitdeluxetable*}{lcccccccccBcccccccBccccc} 
    \setlength\tabcolsep{1.5pt} 
    \tablecaption{\rrr{Galaxy Properties and Emission Line Measurements of \nauroral{} \sagabg{} Galaxies with Secure Auroral Line Metallicities}\label{t:catalog}}
    \tablehead{\colhead{OBJID} & \colhead{RA} & \colhead{Dec} & \colhead{$z_{\rm spec}$} & \colhead{$\log_{10}(\frac{M_\star}{M_\odot})$} & \colhead{$\sigma_{\log_{10}(\frac{M_\star}{M_\odot})}$} & \colhead{$P_{16}[A_V]$} & \colhead{$P_{50}[A_V]$} & \colhead{$P_{84}[A_V]$} & \colhead{$P_{16}[T_e(OIII)]$} & \colhead{$P_{50}[T_e(OIII)]$} & \colhead{$P_{84}[T_e(OIII)]$} & \colhead{$P_{16}[T_e(OII)]$} & \colhead{$P_{50}[T_e(OII)]$} & \colhead{$P_{84}[T_e(OII)]$} & \colhead{$P_{16}[F_{\rm H\alpha}]$} & \colhead{$P_{50}[F_{\rm H\alpha}]$} & \colhead{$P_{84}[F_{\rm H\alpha}]$} & \colhead{$F_{\rm Halpha,lim}$} & \colhead{$P_{16}[\log_{10}(\rm O/H)+12]$} & \colhead{$P_{50}[\log_{10}(\rm O/H)+12]$} & \colhead{$P_{84}[\log_{10}(\rm O/H)+12]$}\\ \colhead{ } & \colhead{$\mathrm{{}^{\circ}}$} & \colhead{$\mathrm{{}^{\circ}}$} & \colhead{ } & \colhead{ } & \colhead{ } & \colhead{$\mathrm{mag}$} & \colhead{$\mathrm{mag}$} & \colhead{$\mathrm{mag}$} & \colhead{$\mathrm{K}$} & \colhead{$\mathrm{K}$} & \colhead{$\mathrm{K}$} & \colhead{$\mathrm{K}$} & \colhead{$\mathrm{K}$} & \colhead{$\mathrm{K}$} & \colhead{$\mathrm{10^{-17}\,erg\,s^{-1}\,cm^{-2}}$} & \colhead{$\mathrm{10^{-17}\,erg\,s^{-1}\,cm^{-2}}$} & \colhead{$\mathrm{10^{-17}\,erg\,s^{-1}\,cm^{-2}}$} & \colhead{$\mathrm{10^{-17}\,erg\,s^{-1}\,cm^{-2}}$} & \colhead{$\mathrm{}$} & \colhead{$\mathrm{}$} & \colhead{$\mathrm{}$}}
    \startdata
    915501850000001608 & 206.09614 & 41.55101 & 0.0370 & 9.14 & 0.2 & 0.0463 & 0.108 & 0.189 & 8870 & 10900 & 12600 & 8820 & 9520 & 10300 & 2210 & 2240 & 2280 & 17.6 & 8.44 & 8.57 & 8.71 \\
    903485560000002676 & 227.09712 & 3.04321 & 0.1644 & 8.64 & 0.2 & 0.0102 & 0.0334 & 0.0758 & 15000 & 16200 & 17400 & 13100 & 14700 & 16200 & 193 & 194 & 195 & 4.13 & 7.83 & 7.92 & 8.03 \\
    916052160000001588 & 174.86955 & 56.12117 & 0.1346 & 8.97 & 0.2 & 0.281 & 0.686 & 1.14 & 12600 & 15400 & 18100 & 11300 & 13600 & 16200 & 154 & 155 & 157 & 5.26 & 7.91 & 8.1 & 8.36 \\
    \enddata
    \tablecomments{\rrr{A full version of this table with all of the emission lines measured for this work is published in its entirety in machine-readable format. An abbreviated version of the table is shown here for guidance. We report the $X^{\rm th}$ percentile of the estimated posterior of quantity $Y$ as $P_{X}[Y]$. For each emission line we also report
    limiting fluxes computed directly as 3 times the standard deviation of the specific flux in the featureless regions of the spectra near each emission line of interest, holding the shape of the line fixed ($F_{\rm lim}=3\sigma_{F_\lambda}\sqrt{2\pi}\ell$, where $\ell$ is the best-fit width of the emission line).}}
\end{splitdeluxetable*}
\section{Modeling the \sagabg{} Sample in $M_\star$-$Z_O$-redshift Space}\label{s:modeling}
\iffirst{
To model the observed evolution of our sample we take a differential approach.
We use the observed low-redshift galaxy sample as a tautologically realistic boundary condition
and attempt to reproduce the 
evolution of the observed sample in stellar mass--metallicity--star formation rate space. 
Given a galaxy's star formation rate (SFR) and stellar mass and some finite timestep $\Delta t$, the model 
removes from the galaxy both the mass in stars and the mass in oxygen that was created over $\Delta t$ given the SFR, and restores to the 
galaxy the gas that was driven out by the star formation that occurred over $\Delta t$ given a mass-loading factor $\etam$.

Starting with our lowest redshift galaxies as a ``reference'' sample, we predict how these galaxies move through 
both SFR--\logmstar{} and $Z_O$--\logmstar{} space with increasing redshift (or, equivalently, lookback time). 
Our model has three fit parameters: a 
parameterization of the galaxies' recent star formation histories ($a_{\rm SFH}$),  \rrr{the mass-loading factor} ($\etam{}$, \rrr{efficiency at which star formation feedback expels gas}), and 21cm-bright gas fraction ($\fhi{}$). 
}
\ifsecond{
We introduced a simple model to reproduce the observed shift in the observed star-forming main sequence 
of the \sagabg{} galaxies in \citetalias{otherpaper}. Here, we will expand that model to include 
a treatment for the galaxies' gas reservoirs and gas-phase oxygen abundances. 

As before, our model 
is differential in nature. That is,
we use our lowest redshift galaxies as our most complete ``reference'' sample and \rrr{retrogress} those 
galaxies to higher redshifts given their observed star formation rates (SFR), an estimate of
their gas masses ($M_g$, which will be discussed in detail in \autoref{s:modeling:reference}), and observed oxygen abundances ($Z_O$ where $Z_O\equiv M_O/M_g$, the ratio between oxygen mass and total gas mass). We then apply the same 
observational constraints to the ``\rrr{retrogressed}'' reference sample as are present in the 
real data to obtain a prediction for our observations at higher redshift.
}

\subsection{Model Assumptions}
\iffirst{
In constructing this differential model we assume the following:
\begin{description}
  \item [Relative completeness] We assume that our low-redshift galaxy reference sample is the most complete subset of our sample. That is, any galaxy at higher redshift would be detectable if it were at the distance of the reference sample. 
  \item [Evolutionary link] We assume that evolving the reference sample backwards in time should, once our detection and selection functions have been applied, reflect the higher redshift samples. That is, we assume that there is an evolutionary link between the reference sample and higher redshift bins. 
  \item [Accretion] We assume that the accretion of stars is negligible compared to in-situ star formation over the last \tlbmax{} and that
  mass inflow balances mass outflow and gas consumption by star formation. 
\end{description}
}
\ifsecond{
This model includes all of the assumptions made in \citetalias{papertwo} and adds extra stipulations to account for the chemical and gaseous evolution of our galaxies. We briefly recount and introduce these assumptions below:
\begin{description}
  \item [Relative completeness] We assume that our low-redshift galaxy reference sample is the most complete subset of our sample, such that any galaxy observed at higher redshift would be detectable if it were at lower redshift.
  \item [Evolutionary link] We assume that \rrr{retrogressing} the reference sample in time should, once our detection and selection functions have been applied, reflect the higher redshift samples. In other words, we assume that we are not imparting a cosmological bias by basing our differential analysis on our lowest redshift sample.
  \item [Accretion] We assume that the accretion of stars and gas is negligible compared to in-situ star formation, winds, and fountain flows that recycle on timescales $<400$ Myr over the last \tlbmax{}. 
\end{description}
}
\iffirst{
  We justify the first assumption as follows: more distant galaxies will have lower \halpha{} fluxes for a given \halpha{} luminosity, and redshift-induced observational reddening in our bands of interest indicates that we will tend to select bluer and more star-forming galaxies at fixed stellar mass and increasing redshift. 
Because \sagabg{} preferentially contains blue background galaxies via the SAGA photometric
selection, the first assumption should be naturally satisfied.

The second assumption should be generally satisfied given our relative completeness assumption and the fact that the rate of stellar mass accretion is small compared to our stellar mass range, though there will be some galaxies that enter or leave the selection area (e.g. stellar mass exceeds $M_\star =10^{10}M_\odot$, quenching). 
Our stellar mass range is selected such that 
self-quenching in the past \tlbmax{} should be a marginal effect --- in the field,
$\sim 5\%$ of galaxies at \logmstar{}$=9.75$ and $<1\%$ of those at \logmstar{}$<8.5$ are quenched \citep{geha2012}. 

We do not make an explicit environmental cut on our sample, and should thus consider whether the divergent evolution of field and satellite galaxies would affect our analysis. Because we select for star-forming galaxies due to our line detection criterion and low-mass, blue galaxies due to our photometric selection, we find that our satellite fraction is both small and constant.
We show in 
\autoref{s:appendix:environment} that our
satellite fraction is consistent with $\approx 5\%$ across all redshift bins.

There should also be galaxies at the high stellar
mass end of our sample that will reach \logmstar{$>10$} by our $z\sim0$ reference sample. However, our comparison should not be affected by these galaxies given that we are performing a relative optimization (and by construction, none of our models will produce galaxies that exceeded \logmstar[$=10$] at high redshift and became \logmstar[$<10$] at $z\approx0$).

Finally, due to a declining stellar-to-halo mass relation over time, 
low-mass galaxies are not generally expected to accrete significant stellar mass via minor mergers \citep{purcell2007, brook2014}. 
Though major mergers between dwarf galaxies do occur, these are relatively rare at low redshift, with
pair and merger searches yielding an expected merger incidence of a few percent \citep{stierwalt2015,besla2018, kadofong2020a}. We thus expect stellar mass build-up due to accretion to be 
negligible in the past \tlbmax{} for our sample on a statistical level. 
}
\ifsecond{
The first two assumptions in our model remain unchanged from \citetalias{otherpaper}.
First, more distant galaxies will have lower \halpha{} fluxes for a given \halpha{} luminosity. The 
observational reddening in the $g$ and $r$ bands will also cause us to 
select intrinsically bluer and more star-forming galaxies at fixed stellar mass and increasing redshift. 
Our spectroscopic targeting probability increases for bluer galaxies and our redshift success rate increases 
for more vigorously star-forming galaxies (due to an increase in \halpha{} luminosity), our first 
assumption should hold true on sample level.

There will be some galaxies that enter or leave the parameter space of selection
(e.g. stellar mass exceeds $M_\star =10^{10}M_\odot$, quenching). 
However, our stellar mass range is selected such that 
self-quenching in the past \tlbmax{} should be a marginal effect \citep{geha2012} and our 
satellite fraction is both small and consistent ($\approx 0.06$) across the redshift range of our sample.
}

In our fiducial model, we will assume that the inflow of pristine gas balances the consumption of gas 
and mass outflow by galactic winds, i.e., that $\dot M_{g} = 0$.
The exact nature of mass inflow into low-redshift galaxies is still a highly open question, especially 
at the stellar mass range of interest in this work \citep{rubin2012, diteodoro2021}. 
Thus, to estimate the effect that this assumption has on our estimates of the mass-loading factor, we run three additional models in which 
we modify our inflow assumption: a model where pristine gas accretion scales with the star formation rate as defined by an additional free parameter such that $\dot M_{\rm in} = \alpha \text{SFR}$ (following~\citealt{schmidt2016} and~\citealt{krumholz2018}), a model where inflow of enriched gas from the CGM balances mass loss to the reservoir, and a model where no accretion is allowed at all. 

Neither our estimate of the mass-loading factor nor our estimate of the average \HI{} fraction in our galaxies are 
shifted beyond their fiducial 68\% confidence intervals; as such, because neither of these accretion models is as well-motivated for our mass regime as our fiducial model, we continue with our assumption of negligible accretion. Further discussion of the inflow models can be found in \autoref{s:appendix:withaccretion}.

{}

\subsection{Sample \rrr{retrogression}}\label{s:modeling:devolution}
We adopt a variation on the classic leaky box model \rrr{wherein outflows are balanced by pristine gas inflow} to 
model the \rrr{retrogression with increasing lookback time} of our reference sample of lowest redshift  ($z<0.035$) galaxies --- the construction of which we discuss in the following section \rrr{---} and assume
that the star formation history of the average galaxy \rrr{in the past \tlbmax{}} can be approximated as follows. 

We write the star formation rate of a given reference sample galaxy at some 
lookback time in terms of the $z<0.035$ SFR and \rrr{one of our model parameters,} $\asfh{}$:

\begin{equation}\label{e:sfh}
  \text{SFR}(t_{lb}) = \text{SFR}(t_{lb}=0)[1+\asfh{}t_{lb}].
\end{equation}
Unlike chemical evolution models, we model (and observe) the star 
formation rate evolution itself rather
than derive a star formation rate while fitting, e.g., a depletion time. This very simple form for the recent average change in the SFR at fixed stellar mass is adopted for two reasons. 
First, with some foresight, we will find that this form for $\text{SFR}(t_{lb})$ is able to describe the observed change in the sample over our redshift range (we return to this point in \autoref{s:results:sfms}). A more complex model for the recent
average star formation history is therefore not motivated by the data.
Second, because we are using real low-redshift galaxies as a reference sample, the processes that drive scatter in the $M_\star-$SFR plane should already be present in our description of this sequence.
These include both physical processes \rrr{such as} the stochasticity of sampling the \rrr{initial mass function and} cluster mass function \rrr{may drive increased scatter between SFR tracers at the \halpha{} luminosities considered} \citep{fumagalli2011}\rrr{,} bursts of star formation on timescales small or comparable to our $\Delta t_{lb}$ \citep{emami2019}, \rrr{as well as} observational effects such as the validity of the constant 10 Myr-averaged SFR assumption in \halpha{} SFR calibrators \citep{kennicutt2012}. 

Given the star formation rate of a galaxy at $t_{lb}=t_i$, we would like to predict how the stellar mass, gas mass, and mass in oxygen has changed from lookback time $t_i$ to a more distant lookback time $t_{i+1}$ where $\Delta t = t_{i+1}-t_i >0$. As noted above, given the mass and redshift range of our sample, we ignore the contribution of accreted stars to the total stellar mass growth in our model. The change in stellar mass can then be written as:
\begin{equation}
  \dot M_\star(t_i) = \text{SFR}(t_i).
\end{equation}\label{e:mstartlb}
Note that because we are going backwards in cosmic time (increasing lookback time), $\Delta t_{lb} < 0$ and thus $\dot M_\star (t_i) \Delta t_{lb} < 0$. 

Similarly, star formation both increases the mass of gas-phase oxygen through stellar production and depletes the mass of oxygen by locking mass in stars and expelling mass through star formation feedback:
\begin{equation}\label{e:motlb}
  \dot M_O(t_i) = (p_O - Z_O(t_i)\left[1 + y_Z \etam{} - f_{\rm rec} \right])\dot M_\star,
\end{equation}
\rrr{which directly invokes our model parameter $\etam{}$, the mass-loading factor.} We assume \rrr{pristine gas accretion,} a stellar yield of $p_O=0.035$ as computed by \cite{vincenzo2016}\footnote{Note that $p_O$ is equivalent to $y_O$ in Equation 4 of \cite{vincenzo2016}. The locked-in fraction is already taken into account their tabulated $y_O$, and so is not additionally multiplied by $(1-f_{\rm rec})$ in \autoref{e:motlb}.} using a \cite{kroupa2001} IMF and the yields of \cite{romano2010}. $p_O$ is sensitive to both the stellar library and IMF assumed: \cite{vincenzo2016} find a range of $0.027<p_O<0.039$ for all combinations of considered stellar yields (those of \citealt{romano2010} and \citealt{nomoto2013}) and Chabrier/Kroupa IMFs.
$Z_O(t_i) \equiv M_O/M_g(t_i)$ is the mass fraction of oxygen at $t_{lb}=t_i$ and $y_Z$ is the over-enrichment of the outflows\footnote{To put $y_Z$ in terms of the metal-loading factor $\eta_Z$ one can write $y_Z \equiv \eta_Z/\etam{} Z_{O,\text{ISM}}^{-1}$}. We assume a fiducial $y_Z=2$ \rrr{following \cite{steinwandel2023_metalloading},} but have verified that the results of our analysis are not affected at the $1\sigma$ level if we were to adopt $1<=y_Z<=3$ (where $y_Z=1$ corresponds to perfect mixing between the ISM and outflows).
$f_{\rm rec}$ is the fraction of stellar mass assumed to be immediately returned to the 
ISM under the instantaneous recycling assumption. The recycling fraction is typically calculated 
by assuming that stars of subsolar mass live forever and that stars of supersolar mass 
recycle instantaneously; for a~\cite{kroupa2001} IMF this yields $f_{\rm rec}\approx0.32$. 
We slightly modify this such that stars with lifetimes shorter than $\Delta t_{lb}$ are assumed
to recycle instantaneously ($M\approx 3.5 M_\odot$), which results in a slightly smaller adopted
$f_{\rm rec}=0.19$.

{}
To arrive at our full predictions we evolve our reference sample, which we will discuss presently, out to $z=0.21$.
\rrr{Our fiducial inference uses a redshift bin size of $\Delta z = 0.035$ ($\Delta t_{lb}\approx 477$ Myr), or 6 bins across our redshift range, but we have rerun our inference with 4 and 8 bins to ensure that our choice of binning does not 
affect our results.}

\subsection{Constructing the reference sample}\label{s:modeling:reference}
We divide our sample into bins out to $z=0.21$ 
(\tlb{}$=2.5$ Gyr). 
Our reference sample is comprised of the galaxies in the first redshift bin, at $0<z<0.035$ --- or, equivalently, $0<\text{\tlb}<480$ Myr. This first bin contains \nreference{} galaxies, \nrefauroral{} of which have a reliable auroral line metallicity.

We measure a mass--metallicity relation based on the $z<0.035$ galaxies that
have auroral line metallicities in order to statistically estimate metallicities for the other $z<0.035$ galaxies that do not have auroral line metallicities. We fit the
same functional form as adopted by~\cite{andrews2013} for our nearby mass--metallicity relation:
\begin{equation}
  \begin{split}
  \log_{10}(\rm O/H) + 12 =& \log_{10}(\rm O/H)_{\rm asm} + 12 \\
  & - \log_{10}\left(1 + \left( \frac{M_{\rm TO}}{M_\star}\right)^\gamma \right),
  \end{split}
\end{equation}
where we find the best-fit parameters using a Gaussian likelihood and Gaussian priors centered on the best-fit values of \cite{andrews2013}. 
We find best-fit values of $\log_{10}(\rm O/H)_{\rm asm} + 12=8.44^{+0.36}_{-0.53}$, 
$\log_{10}(M_{\rm TO}/M_\star) = 8.8^{+1.2}_{-0.93}$, and $\gamma=0.6^{+0.19}_{-0.28}$. We also infer an intrinsic scatter in the 
mass--metallicity relation of ${\sigma_{\rm intr}=0.31^{+0.074}_{-0.086}}$. We find that our low redshift mass--metallicity relation is in good agreement with previous measurements of nearby galaxies, shown using blue \citep{berg2012} and orange \citep{cook2014} points in \autoref{f:localMZR}. \rrr{This agreement indicates that our use of the electron temperature method to determine oxygen abundances has not imposed a significant bias in our sample selection at the lowest redshift bin we consider. }

\begin{figure}[t]
  \centering     
  \includegraphics[width=\linewidth]{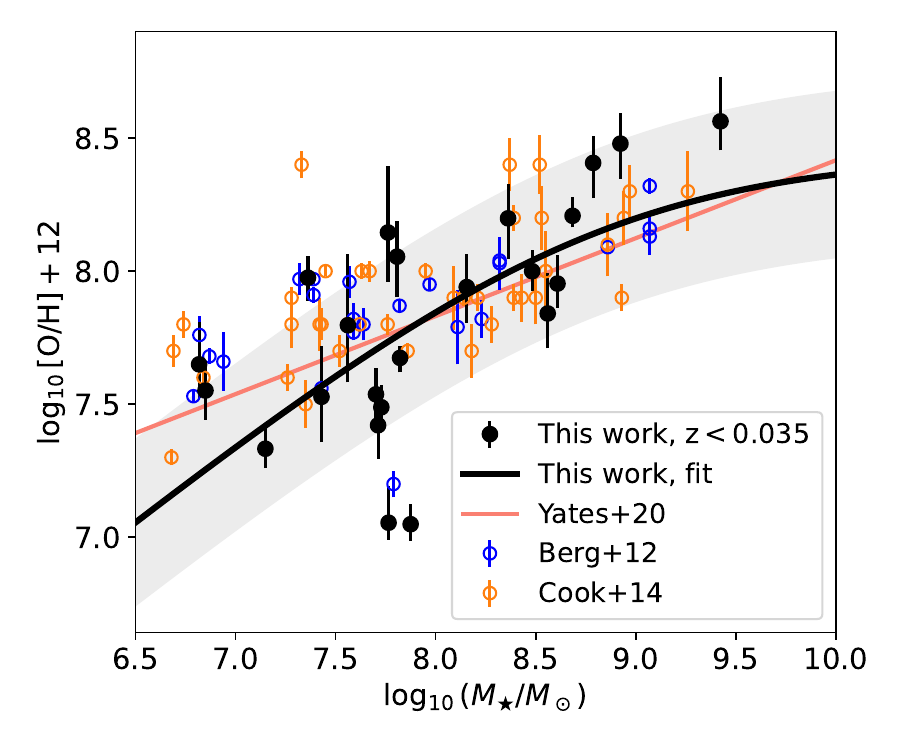}
  \caption{ 
      Stellar mass versus gas-phase metallicity for the galaxies in the reference sample ($z<0.035$)
      for which we are able to measure reliable auroral-line temperature metallicities. Our fit to the $0<z<0.035$ mass--metallicity relation is shown in grey, where the shaded region 
      shows the $1\sigma$ intrinsic scatter of the fit. For context, 
      we also show measurements of nearby galaxy samples using the same temperature-sensitive auroral 
      line ratio approach that we apply here \citep{berg2012, cook2014, yates2020}. All measurements are adjusted to a~\cite{kroupa2001}
      IMF, but otherwise shown as published. We find good 
      agreement between our sample and the literature.
      }\label{f:localMZR}
\end{figure}

Finally, in order to estimate the galaxies' mass in oxygen we need an estimate of the total 
gas mass of the reference sample galaxies.
{}
{}
We adopt individual \HI{} masses for the reference galaxy samples and
fit an average \HI{} gas fraction as a parameter in our model. 28 of our reference galaxies 
have a match in the extragalactic ALFALFA catalog of \cite{haynes2018}; we
estimate an \HI{} mass for the remainder of the galaxies in our reference sample using the \cite{bradford2015} relation.
{}
We find
that the few galaxies that have an ALFALFA detection are in 
good agreement with the estimates from the \cite{bradford2015} relation at their stellar mass. 
In \autoref{s:appendix:gasmass} we take an alternate route to estimating $M_g$ by assuming a fixed 
star formation efficiency, and find that it does not affect 
the outcome of this analysis.

{}

To arrive at an estimate for the total gas mass in our reference sample galaxies we will infer the average mass fraction of 21cm-bright \HI{} as a free parameter in our model:
\begin{equation}\label{e:fhi}
  \fhi{} = \frac{M_{\rm 21}}{M_g},
\end{equation}
where $M_g$ is the total gas mass in hydrogen. This observationally defined quantity primarily traces the 
warm neutral medium (WNM), though there is evidence for significant emission from 
thermally unstable neutral hydrogen in the Solar Neighborhood. The total gas mass in the ISM, however, should include all phases of the ISM; if we assume that it is the warm neutral medium that is traced by 21 cm emission, we would expect a value of $\fhi{}\approx 0.4$ for Solar Neighborhood-like conditions \citep{heiles2003}.

\subsection{\rrr{Comparing Model Predictions to Observations}}\label{s:modeling:observational}

\iffirst{
  Finally, we must also include the effects of our selection function. This includes both the photometric selection described in \autoref{s:sample:photometric}, and the requirement for an emission line detection.
We compute the expected $g-r$ color at luminosity distance $d_{L}(t_{i+1})$ and effective on-sky radius at angular diameter distance $d_{A}(t_{i+1})$ by computing reverse $k$-corrections directly from the spectra; we require that the galaxy have an apparent magnitude of $m_r<21$ \rrr{to} 
remain ``observed'', which we will presently describe, at $t_{i+1}$. \rrr{Galaxies that are flagged as observed at a given lookback time $t_{i+1}$ are excluded from our computation of the likelihood, which we will detail in the following section (\autoref{s:modeling:inference}).}

\rrr{The probability that any given galaxy will be securely redshifted by the SAGA survey is the product of the probability that the galaxy will be targeted, and the probability that the target will be sufficiently bright to obtain a secure redshift. Let us first consider the probability that a galaxy will be targeted.} 

The effective SAGA targeting scheme is based upon the apparent (galactic extinction-corrected) 
$r-$band magnitude of the galaxy, its effective $r-$band surface brightness, and its
$g-r$ color \citep{mao2021}.  
In order to quantify the probability that a \rrr{retrogressed} galaxy would have
been selected for observation, we simply measure the fraction of galaxies that have observations from SDSS, GAMA,
or SAGA AAT/MMT as a function of their (binned) photometric properties. This is done in three dimensions; 
for visualization purposes we show the projected targeted fraction for our photometric parameter space 
in \autoref{f:targeting}. In each panel group, the upper panels show the distribution of all possible 
targets (left, red) and the targeted galaxies (right, blue). 
The main panel shows the fraction of galaxies targeted
as a function of their photometric properties, where
the size of the box corresponds inversely to the uncertainty on the measured fraction assuming a binomial 
distribution. One can clearly see the imprint of the photometric selection described in~\cite{mao2021}, wherein galaxies that are bluer and lower surface brightness are more 
likely to have been observed. 

At each timestep, we compute the observed-frame $r$-band magnitude, $r$-band surface brightness, and $g-r$ color of each \rrr{retrogressed} galaxy. 
\rrr{We compute the expected $g-r$ color at luminosity distance $d_{L}(t_{i+1})$ and the effective on-sky radius at angular diameter distance $d_{A}(t_{i+1})$ by computing reverse k-corrections directly from the spectra. We do not account for the change in the age and metallicity of the underlying stellar populations when computing this shift but note that the change in apparent magnitude due to the change in stellar mass and stellar age is small ($\Delta m_r\sim 0.1$) over our redshift range compared to the effect on the apparent magnitude from distance ($\Delta m_r \lesssim 7$).}

\rrr{At} each timestep we estimate the targeted fraction by drawing from a uniform distribution bounded by the 95\% 
confidence interval. 
We then sample from a uniform distribution bounded by zero and unity, and the galaxy is flagged as observed if the targeted probability exceeds the drawn number. 

\rrr{We then must consider whether the SAGA survey would be able to obtain a secure redshift for the galaxy at the proposed timestep.} We estimate the \halpha{} detection limit as three times the standard deviation of the blank spectrum within $140(1+z_{i+1})$ \AA{} of \halpha{} (we mask \halpha{}, \ionline{N}{2}{6548}, and \ionline{N}{2}{6583}). \rrr{To estimate the expected strength of \halpha{}, we} convert SFR$(t_{i+1})$ to an \halpha{} flux using the inverse of \autoref{e:lha2sfr}, i.e.:
\begin{equation}
  F_{\text{\halpha{}}}(t_{i+1}) = \frac{\text{SFR}(t_{i+1})/[M_\odot\ \rm yr^{-1}]}{5.5\times10^{-42}} \frac{1}{4\pi d_L^2(z_{i+1})}.
\end{equation}
To be flagged as observable, we require that $F_{\text{\halpha{}}}(t_{i+1})$ exceed the 
$3\sigma$ detection limit defined above.

}
\ifsecond{
  \input{./observational_constraints_second.tex}
}

\begin{figure*}[htb]
  \centering     
  \includegraphics[width=\linewidth]{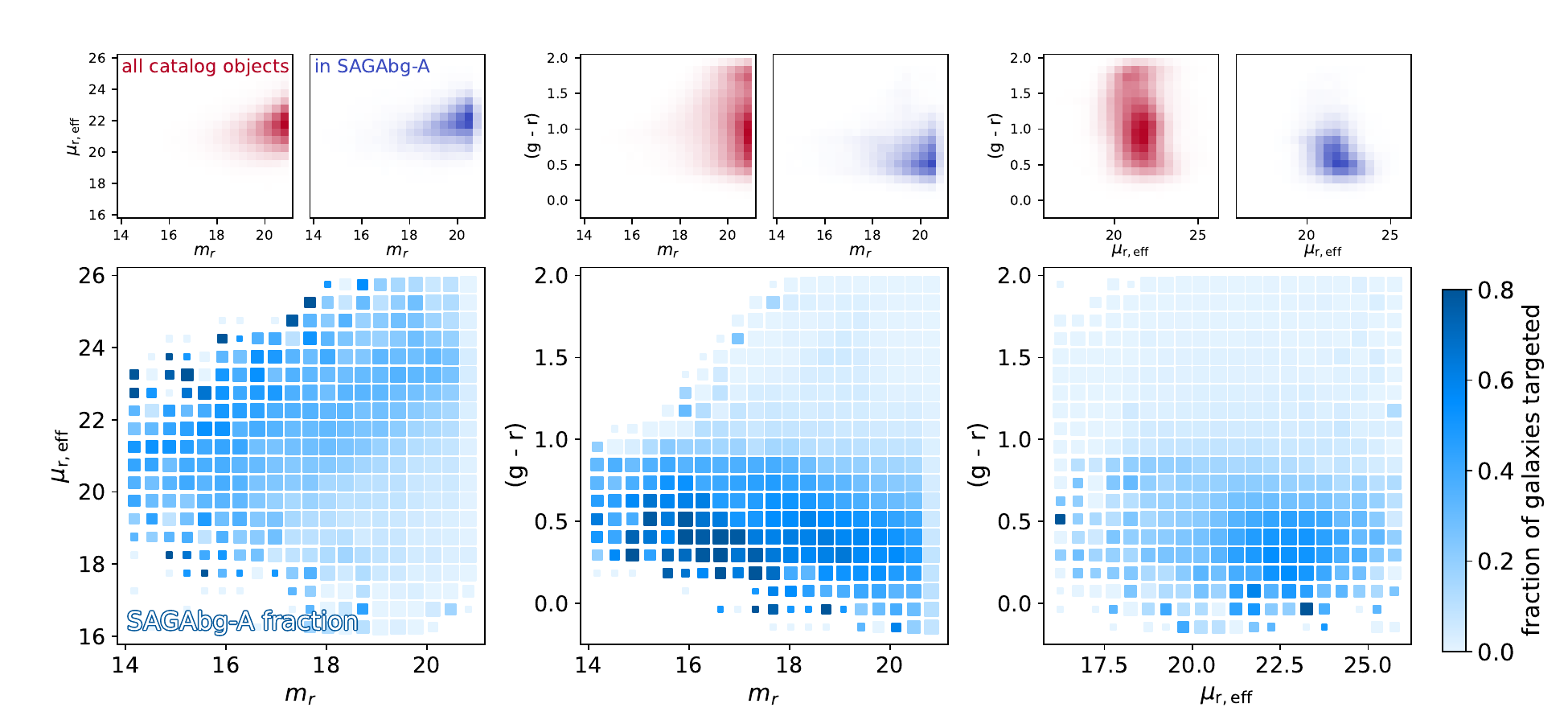}
  \caption{ 
      The effective photometric selection of the present sample. Each panel 
      grouping shows the projected distribution of the targeting catalog (top left, red), the distribution of the present sample (top right, teal), and the fraction 
      of galaxies for a given region of photometric parameter space that are in \sagabg{}. In the main panel, the size of each box corresponds to the inverse of the
      uncertainty on the targeted fraction assuming a binomial distribution.
      }\label{f:targeting}
\end{figure*}

{}

\begin{figure*}[htb]
  \centering     
  \includegraphics[width=\linewidth]{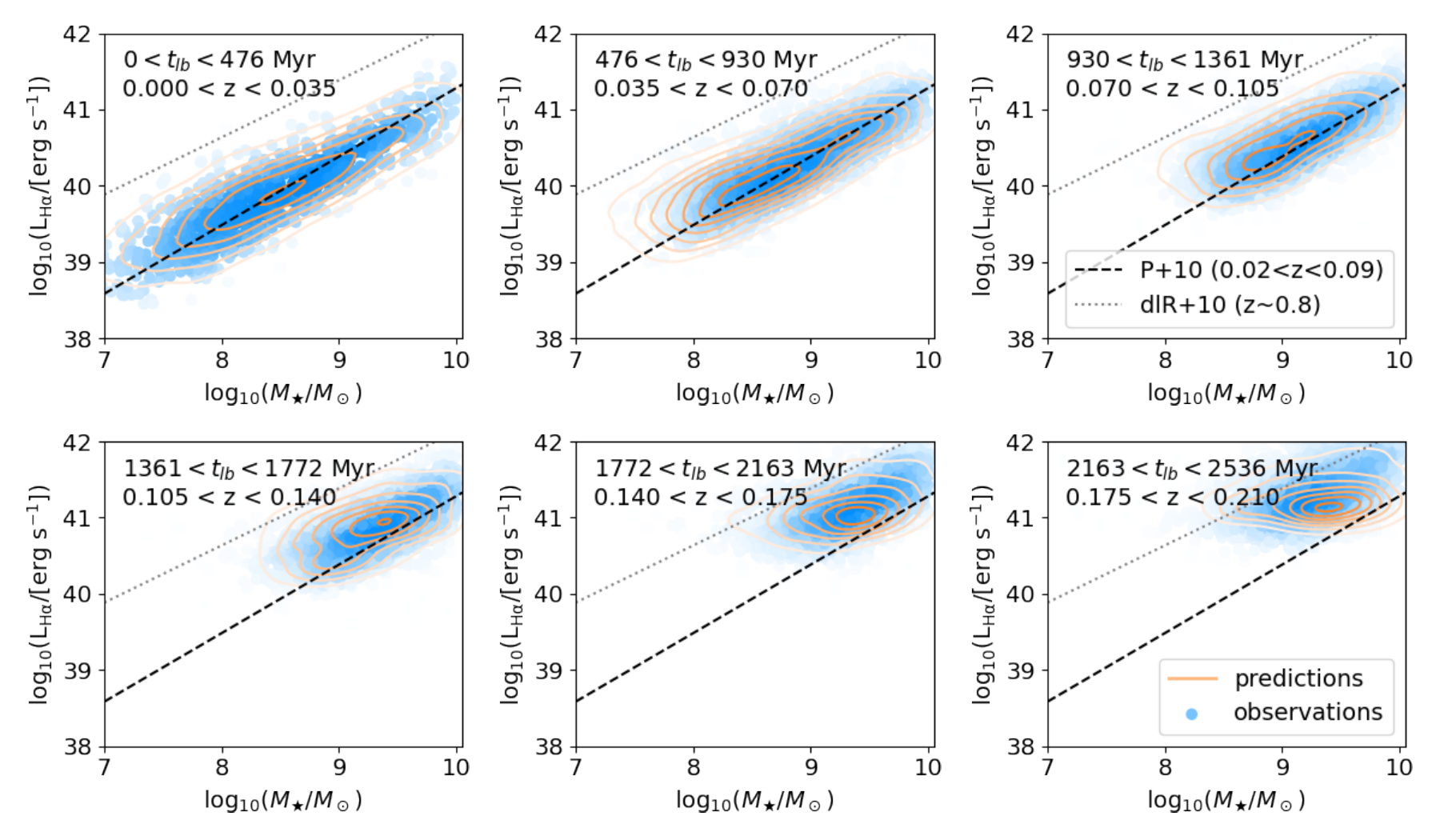}
  \caption{ 
      A comparison between the predicted (orange) and observed (blue) SAGA background sample
      as a function of redshift in the $M_\star$-\lha{} plane. Here we also show
      the SFMS fit of~\cite{peng2010} based off of SDSS ($0.02<z<0.09$) and that of~\cite{delosreyes2015}
      (the narrow-band NewH$\alpha$ survey, $z\sim0.8$) to compare with established literature
      results.
      }\label{f:SFMS_evolution}
\end{figure*}

\begin{figure*}[htb]
  \centering     
  \includegraphics[width=\linewidth]{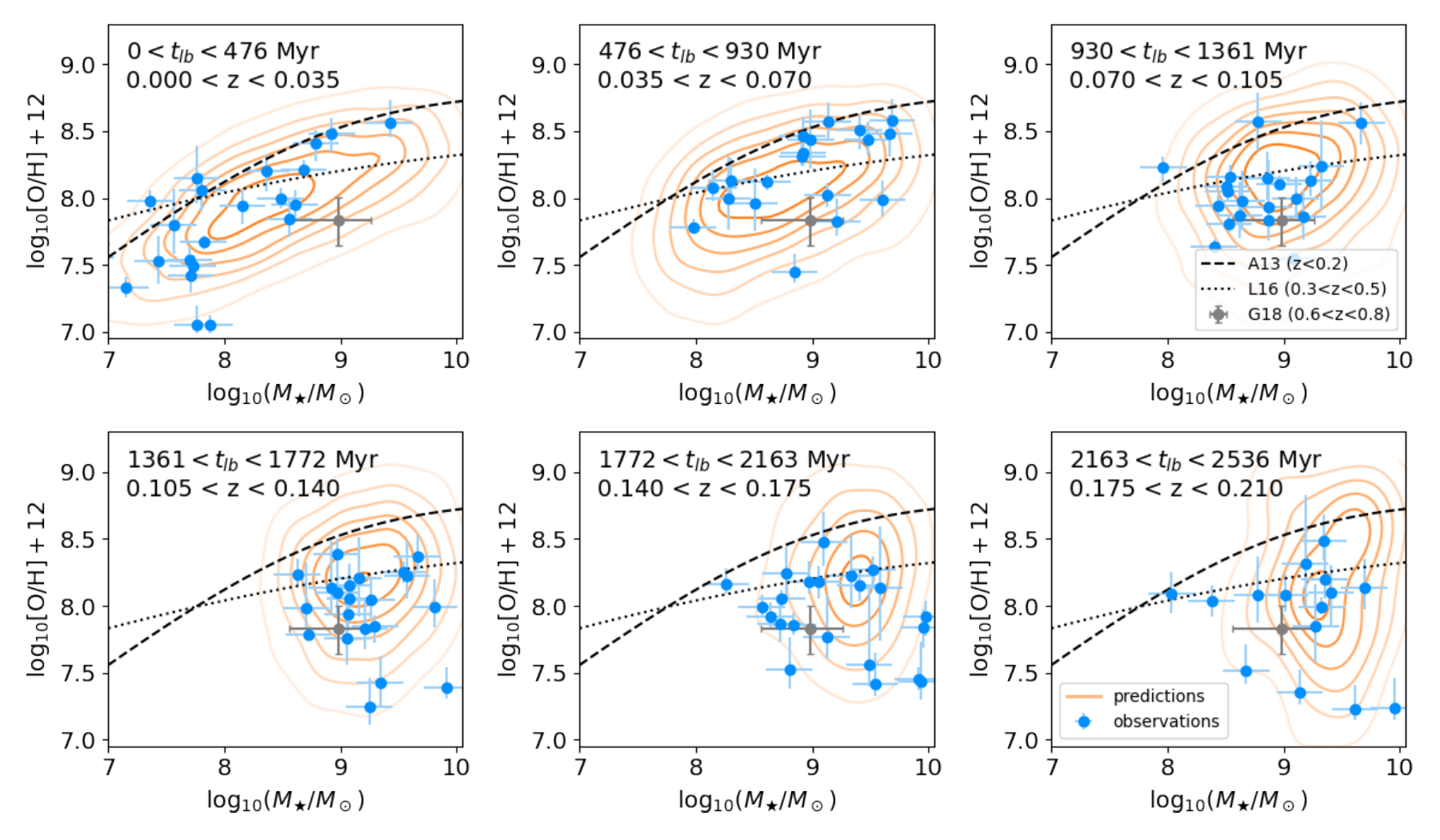}
  \caption{ 
      The same as \autoref{f:SFMS_evolution}, but for the stellar mass-gas phase metallicity plane. 
      Again we show literature results from other studies that derive metallicities from auroral oxygen lines: results from stacked SDSS spectra at $z<0.2$ \citep{andrews2013}, from the 
      MACT survey at $0.3<z<0.5$ \citep{mactii}, and at $0.6<z<0.85$ \citep{gao2018}.
      }\label{f:MZR_evolution}
\end{figure*}

\subsection{Parameter Estimation}\label{s:modeling:inference}
\rrr{With this model of the \sagabg{} selection function in hand, we can now consider how to leverage this information during our inference of the physical parameters of interest}

We \rrr{first} place a uniform prior on all three parameters of interest over a physically reasonable range, where the directly sampled parameter
is $\log_{10}(\etam)$ rather than $\etam$ itself.
\begin{subequations}
  \begin{align}
      \text{Pr}[\asfh{}] &\propto
      \begin{cases}
        \text{const.} & \text{if\ } \asfh{} < 4\\     
        0 & \text{else} 
      \end{cases}\\
      \text{Pr}[\log_{10}\etam] &\propto
      \begin{cases}
        \text{const.} & \text{if\ } 10^{-3}<\etam{}<10^3\\
        0 & \text{else} \\ 
      \end{cases}\\
      \text{Pr}[\fhi] &\propto
      \begin{cases}
        \text{const.} & \text{if\ } 0<\fhi<1\\
        0 & \text{else}. \\ 
      \end{cases}
  \end{align}
\end{subequations}

Here we estimate our likelihood based on density estimation using a 
kernel density estimate over our model predictions for fixed 
parameter choice, sampling from the~\cite{bradford2015}
$M_{\rm HI}-M_\odot$ relation and\rrr{, for reference galaxies that do not have a secure direct temperature method metallicity estimate,} our established reference mass--metallicity relation \rrr{as shown in \autoref{f:localMZR}}. 

As in the ISM property inference, we 
estimate the probability density functions independently and thus the 
joint likelihood can be written as:
\begin{equation}
    \begin{split}
    \ln\mathcal{L}_j(\vec{\text{SFR}},\vec M_\star,\vec Z_O|\vec \theta_m,z_j) &= \sum_i \mathcal{L}_{\rrr{j}}^{\rrr{\rm SM}}(\text{SFR}_i,M_{\star,i}|\vec \theta_{m}, z_j) +\\ 
    &\ \ \mathcal{L}_{\rrr{j}}^\text{\rrr{MM}}(Z_{O,i}, M_{\star,i}|\vec \theta_m,z\rrr{_j})\\
    \ln \mathcal{L}(\vec{\text{SFR}},\vec M_\star,\vec Z_O|\vec \theta_m) &= \sum_j \ln\mathcal{L}_j(\vec{\text{SFR}},\vec M_\star,\vec Z_O|\vec \theta_m,z_j),
    \end{split}
\end{equation}
where $X_i$ indicates the $i^\text{th}$ observation of quantity $X$ \rrr{(SFR, stellar mass, or gas-phase metallicity)} in a given 
redshift bin, $z_j$ is the $j^\text{th}$ redshift bin, \rrr{and $\vec \theta_m$ are the model parameters ($\asfh{}$, $\etam{}$, and $\fhi{}$) adopted at walker step $m$}. \rrr{$\mathcal{L}_j^{\rm MM}$ and $\mathcal{L}_j^{\rm SM}$ refer to the likelihood of the data in metallicity-stellar mass space and SFR-stellar mass space, respectively.}  We estimate the likelihood of the 
higher redshift ($z>0.035$) samples via a Gaussian kernel density estimate \rrr{with bandwidth $N^{-1/(d+4)}$ (i.e. Scott's Rule, \citealt{scott1992}) computed from} the predictions of our model, where we redraw from the $M_\star -M_{21}$ and $M_\star-Z_O$ relations ten times for each density estimate of the predictions. \rrr{In order to incorporate our knowledge of our sample selection function, we exclude retrogressed galaxies whose properties lie outside of the \sagabg{} selection/detection criteria as described in \autoref{s:modeling:devolution} when constructing the density estimate.}

\rrr{This kernel density estimate allows us to arrive at a non-parametric form for both $\mathcal{L}^{\text{SM}}_j(\text{SFR}_i,M_{\star,i}|\vec \theta_{m}, z_j)$ and $\mathcal{L}^{\text{MM}}_j(Z_{O,i}, M_{\star,i}|\vec \theta_m,z_j)$. We then compute the likelihood of observing the sample at the $j$\textsuperscript{th} redshift bin for all galaxies in the \sagabg{} sample at the relevant redshifts. In the case of the star-forming main sequence, we compute the likelihood over all galaxies in the sample. In the case of the mass-metallicity relation, we compute the likelihood over only the galaxies that have a measured oxygen abundance from the auroral lines discussed in \autoref{s:methods:ismconditions}. We note that, as discussed in \autoref{s:appendix:strongline}, our estimate of the mass-loading factor would not be significantly affected if we were to use a strong-line indicator of metallicity that agrees with our low redshift ($z<0.035$) measure of the mass-metallicity relation.}

\rrr{As above, we use the \textsf{emcee} implementation of the Affine Invariant Markov Chain Monte Carlo ensemble sampler \citep{emcee} for parameter inference.}
We run the chain for 10000 steps with 32 walkers and visually confirm chain convergence.

\section{Results}
\subsection{The \sagabg{} Sample in $M_\star$-SFR Space}\label{s:results:sfms}
Here we discuss the observed evolution of the \sagabg{} sample in stellar mass and star formation rate in both our observations and model. The observed evolution is a product 
of both the physical evolution and the effect of our observational limitations and selection on the $M_\star-L_{H\alpha}$ plane as a function of redshift. Becuse the \sagabg{} sample contains only star-forming galaxies (due to our emission line detection requirements), we will compare our results in this section to literature measures of the star-forming main sequence (SFMS).

Each panel in \autoref{f:SFMS_evolution} shows a redshift slice of $\Delta z=0.035$ ($370<\Delta t_{lb}<480$ Myr) with our observed sample shown by the blue points (colored by a Gaussian kernel density estimate to visually indicate density) and our model predictions shown by orange contours. In each panel, we also show the SDSS SFMS ($0.02<z<0.09$) of~\cite{peng2010} as a dashed black line and the New\halpha{} SFMS ($z\sim0.8$) 
of~\cite{delosreyes2015} as a dotted black line; these two surveys are chosen due to their redshift range and use of \halpha{} as a SFR indicator.

We find good agreement between our observations and model predictions both visually in \autoref{f:SFMS_evolution} and
through a quantitative comparison of summary statistics in \autoref{s:appendix:validation}. 
This supports our choice of a simple, linear approximation to the recent SFH of the galaxies in our sample (see \autoref{e:sfh}). The average star formation rate at fixed stellar mass in our sample moves upwards by around 0.4 dex across our sampled redshift range, varying smoothly between the relations of \cite{peng2010} 
 and \cite{delosreyes2015}. The shift of the SFMS shown in \autoref{f:SFMS_evolution}, as stated above, is a combination of a physical shift in the SFMS and an 
observational shift imposed by the \halpha{} flux limit of our spectra\footnote{We trace our detection limits assuming that \halpha{} is the strongest line. There are some cases where \ionline{O}{3}{5007} may be stronger than \halpha{}, but we find that the difference in flux is not large enough to signficantly change the maximum expected distance at which lines may be detected.}. 
In \citetalias{otherpaper} of this series, we \iffirst{will} examine the implications of the physical component of this
shift.

\subsection{The Mass--Metallicity Relation}\label{s:results:mzr}
Having now established that our simple \rrr{retrogression} model reproduces our observed sample in the stellar mass-star formation rate plane, we will now consider the results of the somewhat more complex stellar mass-gas phase metallicity (MZR) plane.

In \autoref{f:MZR_evolution} we show the MZR analog of \autoref{f:SFMS_evolution}, wherein each panel shows our observed auroral line galaxies in blue and model predictions in orange. 
In each panel, we also show literature measurements that use the same metallicity method and bound our redshift limits.
Given the systematic uncertainties that underpin strong line metallicity
calibrations \citep[for a recent review, see][]{kewley2019} and the importance of 
comparing equivalent measures in matters of redshift evolution, we only compare
to literature measurements of metallicity that have been made with the
auroral lines that we consider in this work.
We show mass--metallicity relations derived from stacked SDSS spectra by~\cite
{andrews2013} at $0<z<0.2$, individual galaxy measurements at $0.3<z<0.5$ by~\cite{mactii}, and individual galaxies 
at $1.5<z<3.5$ from the MOSDEF survey by~\cite{sanders2020}. We convert all relations to a Kroupa IMF from a~\cite{chabrier2003} IMF.
At $z<0.035$ our measurements agree well with~\cite{andrews2013}, 
while at $0.17<z<0.21$ our measurements are in good agreement with~\cite{mactii}. 

Similar to the star-forming main sequence, there is good agreement between our observations, model predictions, and literature results. As with the SFMS, we evaluate the success of our model based on the evolution of summary statistics in 
\autoref{s:appendix:validation}, though visual concordance is also shown in \autoref{f:MZR_evolution}.

Previous works have found evidence for an anti-correlation between 
star formation rate and gas-phase metallicity at fixed stellar mass \citep{mannucci2010, andrews2013}, though 
the strength of this effect and its dependence on the method of metallicity estimation is still 
disputed \citep{hughes2013, sanchez2013, telford2016}. Our star formation rate sensitivity limit 
increases with redshift due to distance. To test whether the observed change in the mass--metallicity relation can be entirely attributed to a correlation between SFR and gas-phase metallicity, we consider the galaxies in our sample with star formation rates exceeding the minimum star formation rate observed in our 
sample of \ftst{}-detected galaxies at $z>0.17$ (our highest redshift bin). That is, we consider only the galaxies 
that should have emission lines that are detectable at a distance corresponding to $z=0.17$, all other properties of the galaxy being held constant. We find an anti-correlation between redshift and gas-phase metallicity with 
a best-fit slope of $d(\log_{10}[\text{O/H}])/dz = -1.7^{+0.9}_{-0.7}$, which is consistent with the 
slope of the Taylor expansion of the~\cite{mactii} relation of  $d(\log_{10}[\text{O/H}])/dz =-1.0^{+0.2}_{-0.1}$ 
when measured over $0<z<1$.

Despite using auroral lines to derive oxygen abundances, it is not entirely straightforward to compare our results with that of~\cite{andrews2013}. They 
directly stacked SDSS spectra to measure auroral line metallicities, and the ratio of the mean of the 
emission line need not necessarily be the mean of the emission line ratios (which itself need not reflect the mean of the \telec{} distribution!); nevertheless, it is encouraging to see the agreement with the stacked SDSS results and the present work at $0<z<0.035$.

It is crucial to remember that all of these works 
(the present sample included) are subject to emission line strength biases in that the strong 
line emitters tend to be the galaxies where weak auroral line metallicity measurements may be made. \rrr{However,} the overall body of literature indicates that understanding both observational results \textit{and} observational limitations will be crucial in understanding the history of low-mass galaxy enrichment.

\begin{figure*}[t]
  \centering     
  \includegraphics[width=0.6\linewidth]{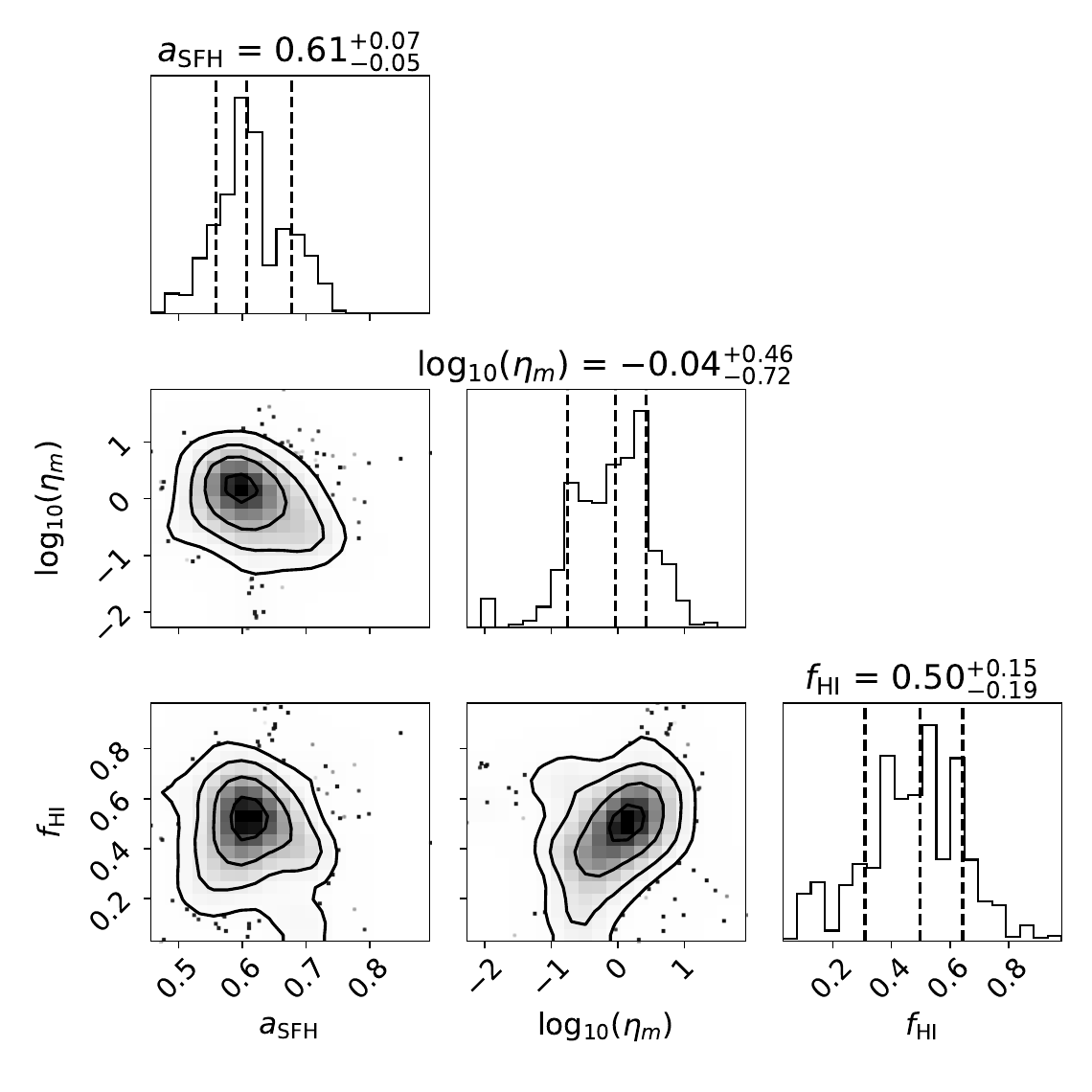}
  \caption{ 
      Posterior distributions of our modeled parameters $\asfh{}$, 
      $\etam{}$, and $\fhi{}$. The posterior distributions shown here are smoothed by a 
      Gaussian kernel of $\sigma =1.5$.  
      }\label{f:corner}
\end{figure*}

\subsection{Model Fits: $\etam{}$ (and $\asfh{}$, $\fhi{}$)}\label{s:results:parameters}
Having now demonstrated that our model is able to reproduce the observed \sagabg{} sample distribution in $M_\star$-SFR-$Z_O$-$z$ space, let us consider the inferred parameters themselves. As laid out in \autoref{s:modeling}, we fit three parameters in this model: $\asfh{}$, which 
parameterizes the average increase in SFR as a function of redshift (\autoref{e:sfh}), 
$\etam{}$, the mass-loading factor (\autoref{e:motlb}), and $\fhi{}$, the
average 21cm-bright \HI{} mass fraction (\autoref{e:fhi}).
In \autoref{f:corner} we show a corner plot of the inferred parameters $\asfh{}$, $\etam{}$, and $\fhi{}$. 
There is no evidence for strong correlations in the joint posteriors of our parameter inference, and we report the median and 68\% scatter in \autoref{f:corner}.

Taking the inferred parameters in turn, let us first consider the recent evolution in the average SFH of the sample. We find evidence for a moderate increase in SFR as a function of lookback time, indicating that the shift in the observed SFMS cannot be fully explained by our observational limits, and is due in part to a physical shift upwards in the star-forming main sequence from \tlb{}$\sim 0$ to \tlb{}\tlbmax{}.
This effect will be more thoroughly examined in the second paper of this 
series \citeprep{Kado-Fong}.

Finally, we arrive at an estimated mass-loading factor of \etamest{}. This inferred 
value for the mass-loading factor is close to unity (i.e., one solar mass of gas expelled per year for each solar mass of stars formed), and was derived using a framework that differs significantly from direct measures of the mass-loading factor. In the following discussion, we will consider the implications of this relatively low mass-loading factor in the broader context 
of previous observational measures of $\etam{}$ and current theoretical predictions for mass-loading in low-mass galaxies.

Because $\etam{}$ could be a mass-dependent quantity, it is also important to constrain the stellar mass range over which this $\etam{}$ inference is sensitive. To do so, we rerun our inference while varying both the upper and lower limits on stellar mass until the uncertainty on any of the inferred parameters exceeds twice that of the uncertainty in the fiducial run. The effective stellar mass range of the sample as characterized by this empirical method is $10^{7.5}\lesssim M_\star/M_\odot \lesssim 10^{9.5}$. This range also corresponds roughly to the
10\tth{} and 90\tth{} percentiles of the stellar mass distribution of the reference sample. 
{} 

{}

\section{Discussion}
We now consider the implications of our estimated mass-loading factor on the 
contemporary landscape of measurements and predictions of the mass-loading factor.
This section begins with a comparison between our inferred $\etam{}$ and observational results from the 
literature that are based on direct measurements of mass outflow rates (\autoref{s:discussion:direct}). We then compare to 
contemporary results from models and simulations of low-mass galaxies in \autoref{s:discussion:simulations}
before touching on the limitations of our simple model for the low-mass galaxy population in 
\autoref{s:discussion:limitations}.

\begin{figure}[htb]
  \centering     
  \includegraphics[width=\linewidth]{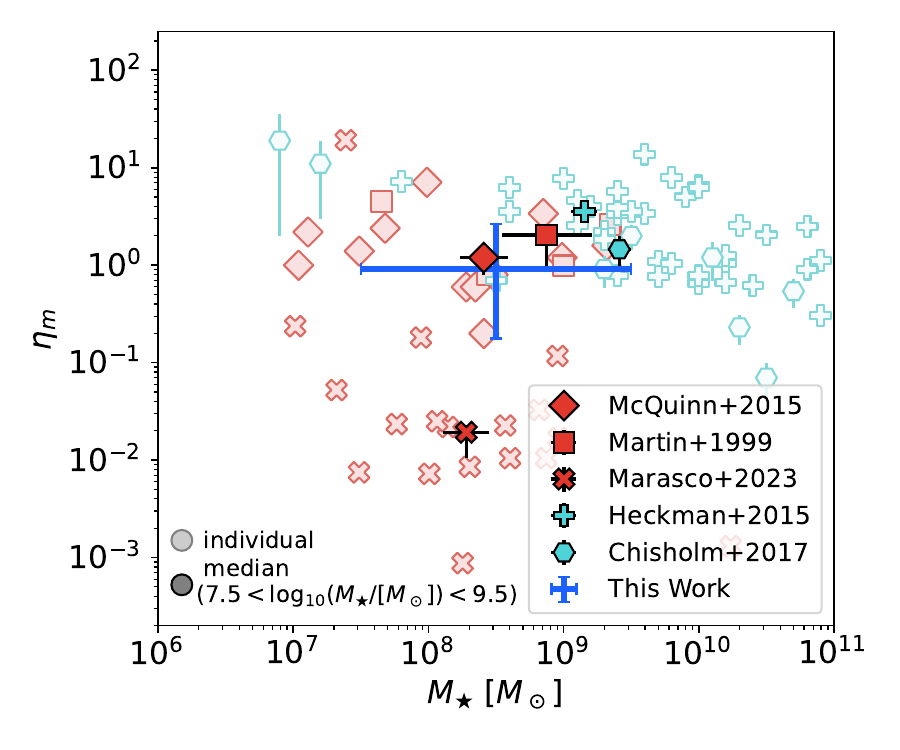}
  \caption{ 
    Mass loading factor vs. stellar mass, comparing our results to direct observations of outflowing gas. These studies either rely on detection of outflowing gas in \halpha{} emission (red circles show \citealt{mcquinn2019}, red 
    squares show \citealt{martin1999}; red line shows \citealt{marasco2023}) or from UV absorption features (aqua crosses show 
    \citealt{heckman2015}; aqua circles show \citealt{chisholm2017}).
    For each study, we show both the individual galaxy measurements as light points and the median of the sample in the 
    stellar mass range $7.5<\log_{10}(M_\star/[M_\odot])<9.5$
    as the dark points with black outlines. 
    The uncertainties on the median of the literature samples are estimated via bootstrap resampling. 
    }\label{f:massloading_direct}
\end{figure}

\subsection{Comparison to Direct Measures of $\etam{}$}\label{s:discussion:direct}
There are two main methods of measuring $\etam{}$: direct and indirect. Direct measures of $\etam{}$ are typically made for individual galaxies by tracing gas that has been deemed to be outflowing by some definition, while indirect measures trace the impact that outflows 
enact on some other aspect of galaxy evolution at a sample or population level. The study at hand is a type of indirect observation, albeit one done in a differential manner.

Our value of $\etam{}$ is in good agreement with most direct measurements of galaxies in the relevant stellar mass range. In \autoref{f:massloading_direct}, the marker shape indicates the 
individual study while the color represents the type of measurement: red points indicate direct measurements via \halpha{} 
emission, while turquoise points indicate direct measurements via UV absorption.

We find good agreement between our results and those of~\cite{mcquinn2019} and~\cite{martin1999}, both of whom use 
\halpha{} emission to identify and measure mass outflow rates. 
The spectroscopic study of~\cite{marasco2023} find a near-negligible mass-loading factor of $\etam{}\approx0.02$ for low-mass 
galaxies based off of an analysis of the flux-weighted velocity distribution of \halpha{} lines in integral field unit (IFU) spectroscopy. A discrepancy between the literature results may 
be related to a difference in methodology; whereas the first two studies identify \halpha{} emitting gas that is spatially offset from host dwarf galaxies \citep{martin1998, martin1999, mcquinn2019}, \cite{marasco2023} decompose the full \halpha{} line profile of the galaxy of face-on galaxies into outflowing and non-outflowing components.

A mass-loading estimate based on \halpha{} should
trace only $\etam{}$ of gas in the warm phase, whereas the present work does not distinguish between ISM phases. The concordance is not surprising, however, if we consider that the warm phase is predicted to dominate mass outflow \citep{kim2020,steinwandel2022b}. 
An estimate of $\etam{}$ as defined by \autoref{e:motlb} could therefore be reasonably expected to be in agreement with measures of $\eta_{\rm m,warm}$, the mass-loading factor in the warm (ionized) phase.

We derive an estimate for the mass-loading factor that is somewhat lower than the estimates that have been made from 
UV absorption features. 
In the overlapping region of stellar mass space, we find good agreement with~\cite{chisholm2017}, though we caution that 
this includes only two galaxies from their sample. The
\cite{heckman2015} median mass-loading factor, $\etam\approx2.5$, is slightly higher than our estimate. 
Both UV-based studies 
shown here have a high fraction of starbursting dwarfs; a change in $\etam{}$ as a function of 
SFR or $\Sigma_{\rm SFR}$ is another plausible explanation for the offset in $\etam{}$ given the
spatially coherent nature of star formation and star formation-driven outflows. 

We note that is also not straightforward to compare \halpha{} and UV-based studies against each other.
First, the \halpha{} literature studies and this work use \halpha{}-derived SFRs, which probe $\sim 5$ Myr timescales, while UV-based star formation rates 
probe significantly longer ($\sim 100$ Myr) timescales. Our own indirect estimate of $\etam{}$ 
is averaged over $\sim 400$ Myr given the size of our redshift bins. Thus, we
may expect our estimate of $\etam{}$ to be biased low compared to direct measurements if the 
majority of the outflowing mass recycles on timescales short compared to 400 Myr. In practice, 
the concordance between our results and the direct measurements presented here implies that
this effect does not strongly impact the comparison. Indeed, the study most significantly offset
from our results and other literature results \citep{marasco2023} estimates 
mass-loading factors significantly \textit{lower} than our own estimate.

The concordance between the mass-loading factor we present here and the results from direct detection studies is encouraging because these methods to estimate $\etam{}$ demand different simplifying assumptions. Directly measuring outflow rates requires strong assumptions about the geometry and velocity of the outflows, while our differential method requires strong assumptions about ISM mixing and accretion timescales. There is minimal overlap between the assumptions made by direct outflow measures and our indirect differential approach: arriving at the same answer 
is an important measure of observational consensus-building in an arena where theoretical models differ by orders of magnitude.

\begin{figure}[t]
  \centering     
  \includegraphics[width=\linewidth]{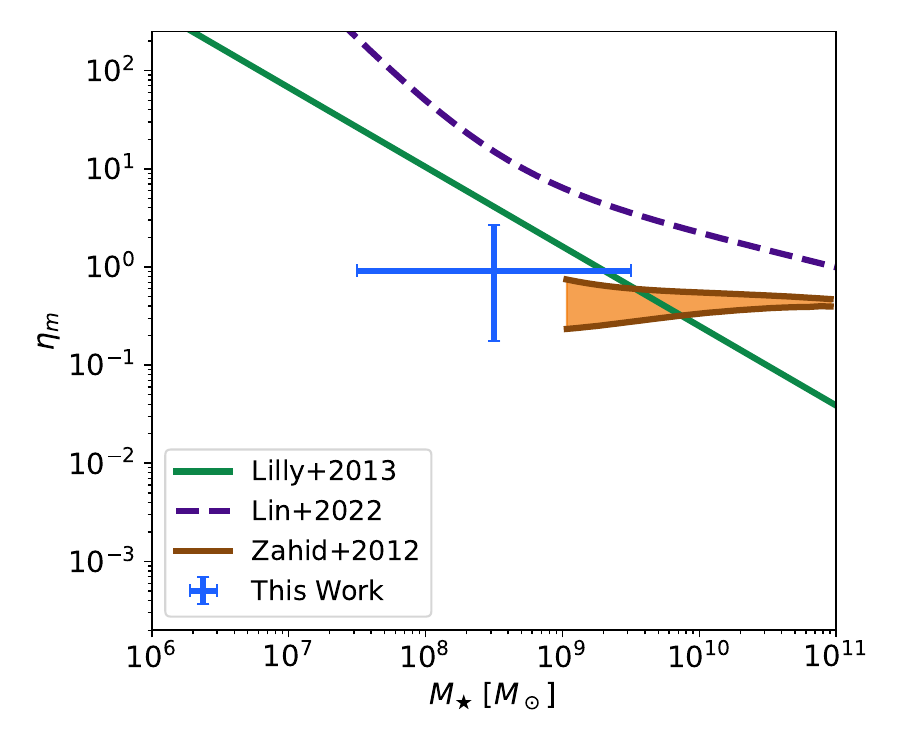}
  \caption{ 
    Mass loading factor vs. stellar mass, comparing our results to models that infer $\etam{}$ as a parameter influencing some other property of the galaxy population. We show the gas regulator model of \cite{lilly2013} as a solid green line, the fit to the mass--metallicity relation of \cite{lin2022} as a dashed purple curve, and the model of the total oxygen galaxy budget of \cite{zahid2012} as an orange shaded region.  
    }\label{f:massloading_indirect}
\end{figure}

\subsubsection{Comparison to $\etam{}$ Inferred from Analytic Models}\label{s:discussion:indirect}
We compare our results to three indirect constraints on the mass-loading factor in low-mass galaxies: two studies that inferred $\etam{}$ from the 
$z\sim0$ mass--metallicity relation (\citet{lilly2013} and \citet{lin2022}), one that compared the mass--metallicity relation between 
surveys at different redshifts \citep{zahid2012}, and one that inferred $\etam{}$ from the stellar-to-halo mass and ISM-to-stellar mass relations \citep{carr2022}.

Our estimate of $\etam{}$ is in good agreement with the gas regulator model of~\cite{lilly2013} at the high mass end of our sample and significantly lower than the chemical evolution model estimate of~\cite{lin2022}, despite the fact that both studies aim to reproduce the observed SDSS $M_\star$-metallicity-SFR distribution at $z\sim0$ to estimate $\etam{}$. We also find good agreement with the inter-survey comparison of SDSS and DEEP2 by \cite{zahid2012}.
All three of these studies use gas-phase metallicities estimated through
strong-line calibrations --- \cite{lilly2013} and \cite{lin2022} moreover use the same SDSS metallicities as measured by \cite{mannucci2010}. These calibrations are known to incur systematic uncertainties of up to $\sim\!1\text{ dex}$ 
both between different line ratio estimators \citep{kewley2008} and between different calibrations of the same line ratio estimators \citep{kewley2019}. A systematic offset in metallicity could explain a difference between our results and those of~\cite{lin2022}, 
but not the difference between
the results of~\cite{lin2022} and~\cite{lilly2013}.  

We also find good agreement with \cite{carr2022} in that they argue for winds with a low mass-loading factor ($0.1\lesssim\etam{}\lesssim 10$) and a high energy loading factor. However, we note that the form of their mass-loading factor dependence with halo mass was chosen to agree with direct measurements of $\etam{}$, and thus is not independent of the concordance that we find with direct observations of $\etam{}$.

{}

\subsection{Comparison to Theoretical Predictions}\label{s:discussion:simulations}
In \autoref{f:massloading_comparison} we compare to predictions of $\etam{}$ from the simulation literature.
Simulations provide a more direct view of the mass-loading factor, as outflowing gas can be explicitly traced and tabulated. However, the exact definition of outflowing gas varies significantly from work to work: a prediction for the mass-loading factor is typically set by computing mass flux for some subset of the gas classified as outflowing through a slab or shell displaced a few kpc from the galaxy of interest. Outflows are either
selected to be all gas that is moving away from the midplane \citep[$v_z>0$ or $v_R>0$, depending on the coordinate system used, see e.g.][]{muratov2015,kim2018,hu2019,nelson2019,steinwandel2022a,steinwandel2022b} or a subset of the gas that is predicted to escape to a given midplane distance \citep{anglesalcazar2017, pandya2021}.
Both the height and outflow selection criteria can change the estimate of $\etam{}$ by a factor of several \citep{nelson2019, pandya2021}; we will discuss the effect of these choices on our approach to simulation comparison below. 

To illustrate the implications of where and how simulators measure their mass-loading
factors, in \autoref{f:massloading_comparison} we show with outlined points only those mass-loading 
factor predictions for a set of works in which the mass outflow rate is computed 
with a velocity cut of $v_{\rm out}>0$. The differences between the predictions here 
should be roughly indicative of the differences in physical prescriptions and numerical effects for these simulations. 
In the same color, we also show an expanded view of the
theoretical landscape which includes a wider range of methods to determine $\dot M_{\rm out}$. 
It is generally understood that differences in 
subgrid models for star formation -- and in particular supernova -- feedback as well 
as numerical methods for how energy and momentum from star formation feedback 
contribute significantly to the range of mass-loading factors seen in simulations.
Furthermore, here we see that $\etam$ can vary 
by a factor of several for the same simulation;
\cite{pandya2021} showed in particular that mass-loading factors derived from the FIRE simulations can vary between $\etam\sim5$ and $\etam\sim30$ for galaxies at 
\logmstar[$\approx 8.5$] due only to differences in how outflows are identified.

\subsubsection{Comparison to High Resolution Galaxy Simulations}\label{s:discussion:simulations:ism}
We consider two resolved-ISM, non-cosmological galaxy simulations close to our mass range: the LMC-mass
galaxy simulation of~\cite{steinwandel2022b}, with a gas mass resolution of $m_{\rm gas}=4M_\odot$, and the lower mass ($M_\star\approx 2\times 10^7 M_\star$) simulation of~\cite{hu2019}, with a mass resolution of $m_{\rm gas}=1 M_\odot$. We 
find that these two simulations, which bound the stellar mass range we probe,
also bound our prediction for $\etam{}$.

Let us first compare to the \cite{steinwandel2022b} of the LMC-mass galaxy. We show the time-averaged
estimate of $\etam{}$ at $z=\{3,5,10\}$~kpc from the disk (where 10 kpc is approximately 0.1 $R_{vir}$). 
There is 
relatively little change in $\etam{}$ as a function of height off the midplane beyond $z\sim 1$~kpc relative to the range of $\etam{}$ proposed by the full set of simulations we consider. We also 
compare to the lower mass simulation of~\cite{hu2019} using their asymptotic value of $\etam{}(R)$. The mass-loading factor estimated by~\cite{hu2019} is somewhat higher than our estimated $\etam{}$. This could be consistent with 
the picture of a stellar mass-dependent $\etam{}$, but as we discussed in \autoref{s:discussion:direct}, it remains 
unclear how strong this mass dependence is in observations.

\srr{We also make a comparison to galactic wind framework developed from the Three-phase Interstellar Medium in Galaxies Resolving Evolution with Star Formation and Supernova Feedback (TIGRESS) simulations \citep{kim2017, kim2020b}. These simulations consider a slice through the galactic midplane at parsec-level resolution; though they span a range of star-forming environments, due to the box size they do not explicitly consider the total stellar or halo mass of the host galaxy. To make a comparison to their results, we use the \textsf{twind} package \citep{kim2020, kim2020b} to measure the range of mass-loading factors expected from TIGRESS when outflows are defined as the gas with Bernoulli velocities that exceed a chosen escape velocity. We adopt three escape velocities: $v_{\rm esc} = \{ 50, 70, 90\}\ \kms{}$, which span the approximate stellar mass range of our sample based off of the spread in the relation between observed stellar mass and halo virial velocity presented by \cite{behroozi2019}. In \autoref{f:massloading_comparison} we show the range of mass-loading factor predicted by the TIGRESS simulations, where the range of the shaded region along the y-axis shows the 
span between the 10$^{\rm th}$ and 90$^{\rm th}$ percentiles computed from \textsf{twind} at $10^{-5}<\Sigma_{\rm SFR}<10^{-1}\ M_\odot\ \rm yr^{-1}\ kpc^{-2}$, and the shaded region along the x-axis
shows the span between the same percentiles based off of the stellar-to-halo relation of \cite{behroozi2019}. We note that although this is an approximate placement of the TIGRESS predictions along
the x-axis due to the uncertainty in the stellar-to-halo mass relation at these stellar masses and the simplistic choice to adopt a single escape velocity, the predicted mass-loading factors are in 
good agreement with our results and much lower than the cosmological simulations (orange and mauve curves), which we will address presently.}

\begin{figure*}[t]
  \centering     
  \includegraphics[width=.83\linewidth]{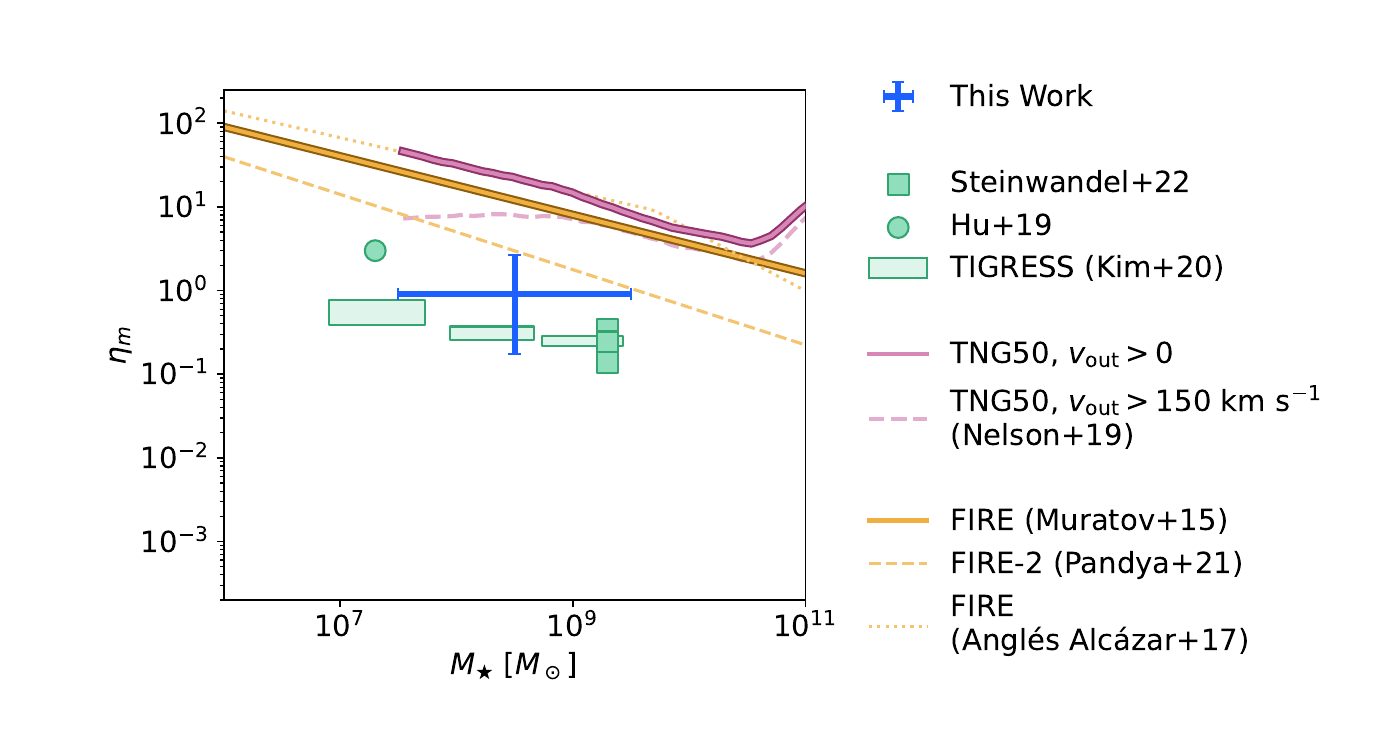}
  \caption{ 
      A comparison between our results (the blue error bar in each panel) to various theoretical predictions from the literature. The color denotes the class of simulation, while the shape of the marker denotes the individual study as described in the text.       
      Individual galaxy simulations are shown in green (the green triangle shows \citealt{hu2019}; the green circles show \citealt{steinwandel2022a} and \citealt{steinwandel2022b}\srr{; the green shaded regions show the expected $\etam{}$ from the TIGRESS simulations based off the mass-loading for outflows with $v_{B,z}>v_{\rm esc}$ where $v_{\rm esc} = \{ 50,70,90\}\kms{}$ and $10^{-5}<\Sigma_{\rm SFR}<10^{-1}\ M_\odot {\rm yr}^{-1} {\rm kpc}^{-2}$. We use the stellar-to-halo mass relation of \cite{behroozi2019} to convert escape velocities to rough ranges in stellar mass.}), 
      the FIRE zoom simulations in orange (the solid orange line shows~\citealt{muratov2015}; the dotted orange line shows~\citealt{anglesalcazar2017}; the dashed orange line shows~\citealt{pandya2021}) and mauve shows results from Illustris TNG50 \citep{nelson2019} where outflows are
      defined as $v_r>0$ (upper curve) and where outflows are defined as $v_r>150\ \kms{}$ (lower curve). All literature references where outflows are defined with a $v_z>0$ cut at some midplane distance are shown by outlined markers.
      }\label{f:massloading_comparison}
\end{figure*}

\subsubsection{Comparison to Cosmological Simulations}\label{s:discussion:simulations:cosmo}
Finally, we consider the big box and zoom cosmological simulations.
All three orange lines in \autoref{f:massloading_comparison} originate from
analyses of the FIRE-1 and FIRE-2 simulations \citep{muratov2015,anglesalcazar2017, pandya2021}, with baryonic mass resolutions of $250\leq m_{\rm bary}\leq 7100 M_\odot$. As demonstrated in~\cite{pandya2021}, the difference
in the three FIRE $\etam{}$ estimates originates from a difference in how 
$\dot M_{\rm out}$ is defined, rather than a difference in physical prescriptions between FIRE-1 and FIRE-2. 
In particular,~\cite{muratov2015} uses a
$v_r>0$ cut at $R=0.25R_{\rm vir}$ (similar to the isolated galaxy simulations),~\cite{anglesalcazar2017} directly tracks the displacement of gas out of the 
galactic midplane, and~\cite{pandya2021} leverages a cut on the 
Bernoulli velocity at $0.1-0.2R_{\rm vir}$
to distinguish between escaping winds and gas that may remain bound at large radii.\footnote{\cite{pandya2021} measures the radial component of the total Bernoulli velocity, which tracks the total specific energy of a particle, to select gas that has sufficient energy to move from some starting galactocentric radius to some larger galactocentric radius; see their Section 3.1.1 for details.} 

Having laid out the guides by which these three works compute $\etam{}$, it is likely that a true comparison to our $\etam{}$ estimate would be somewhere between 
that of~\cite{muratov2015} and~\cite{pandya2021}. The mass-loading estimate 
of~\cite{anglesalcazar2017} likely includes gas that is quickly recycled into the ISM and would therefore be systematically higher than our $\etam{}$ estimate as a matter of definition.
On the other hand, the mass-loading factor estimate of~\cite{pandya2021} could be 
systematically offset to lower $\etam{}$ than our estimate because their definition 
attempts to only capture gas that will truly escape the galaxy, excluding gas that may remain bound at large radii. Because that bound gas would remain unavailable for future generations of star formation (for some timescale that is long compared to our sample redshift range), it should be included in our estimate of $\etam{}$. The positive radial velocity at $0.25R_{\rm vir}$ criterion
that~\cite{muratov2015} make to define $\etam{}$ is also closely aligned to that 
of~\cite{steinwandel2022b}. 

We additionally note that low-mass FIRE galaxies have been found to have low \HI{}-to-stellar mass ratios compared to observed galaxies \citep{elbadry2018,kadofong2022a}; this finding is in agreement with a picture in which the FIRE-1 and FIRE-2 mass-loading factors are systematically displacing more gas from low-mass halos than proceeds in
the observed Universe, though it is certainly not the only explanation for such a discrepancy. 


Few big box cosmological simulations have a sufficient mass resolution to 
interpret the implication of mass-loading on low-mass halos, and even fewer have reported mass-loading factors. With a baryonic mass resolution of $m_{\rm bary}=8\times10^4 M_\odot$, 
Illustris TNG50 \citep{nelson2019} is the highest resolution and smallest box size run of the TNG suite and has published mass-loading factors, though 
these mass-loading factor predictions were computed at $z=2$. 
It is unclear how the mass-loading factor is expected to evolve between $z=2$ and $z=0$, though some simulations have reported negligible
evolution \citep{pandya2021}. We show the TNG50 results as thick mauve curves: the upper curve shows $\etam{}$ at $R=10$~kpc where outflows
are defined by $v_r>0$, and the lower curve shows the same where outflows are defined as $v_r>150~\kms{}$. In both cases the resulting $\etam{}$ 
is higher than our estimate by around an order of magnitude.

\subsection{Limitations and Possible Biases}\label{s:discussion:limitations}
As we emphasized earlier in this text, the model presented in this work is by design a narrow and incomplete view of galaxy evolution --- our present scope is not to instantiate the galaxy population from first (or some facsimile for first) principles but to explain the evolution over a small range of physically interesting parameter space.
{}

In order to reconcile our results with the mass-loading factors produced in cosmological simulations, we would need to be significantly underestimating 
the mass-loading factor. In our fiducial equilibrium model (\autoref{e:motlb}), the derivative of $\dot Z_O$ (the time derivative of the oxygen abundance) with respect to the mass-loading factor can be written as 
\begin{equation}
  \begin{split}
    \frac{d\dot Z_O}{d\etam}(t) =- \frac{Z_O(t) y_z \dot M_\star(t)}{M_g}.
  \end{split}
\end{equation}
If we assume that $\dot Z_O$ is non-negative, 
an increase in the mass-loading factor results in a lower value of $|\dot Z_O(t)|$; that is, a shallower evolution 
in the mass--metallicity relation. If the true mass-loading factor were higher, as cosmological simulations require, we would need to be incomplete in 
low-metallicity galaxies at low redshift and/or incomplete in high-metallicity galaxies at high redshift,
particularly at the high mass end. We find it unlikely that we are incomplete at low metallicity
at low redshift given the concordance of our low-redshift mass--metallicity relation with 
that of deeper nearby galaxy samples (as shown in \autoref{f:localMZR}). \rrr{Though higher metallicity objects tend to have weaker \ionline{O}{3}{4363} lines with respect to the brighter \ion{O}{3} emission lines, potentially imposing a bias at high redshift, the concordance that we reach between strong line and auroral line indicators (see \autoref{s:appendix:strongline}) that it is} 
unlikely that the discrepancy between 
our estimate and theoretical requirements can be explained by observational incompleteness. 

We note that our emission line-based selection produces a sample 
that contains relatively fewer satellite galaxies with respect to the overall 
satellite fraction \rrr{expected} at our stellar mass range (see \autoref{s:appendix:environment} for 
details on our estimated satellite fractions). The mass-loading factors calculated from the high-resolution, non-cosmological simulations are based on completely isolated galaxies, while the mass-loading factors for the cosmological simulations include satellite galaxies. If our concordance with the resolved-ISM simulations were due to 
the impact of environment rather than star formation feedback prescriptions, the lowest-SFR galaxies that are missing from our sample -- at this mass range, the transitional satellites -- would need to have tremendously high mass-loading factors. We consider this situation unlikely: empirical studies have suggested that $\etam$ is either insensitive to or increases with SFR (as discussed in \autoref{s:discussion:direct}). 

Finally, it is also worth considering that the subset of galaxies at a given redshift for which we are able to measure electron temperatures
via auroral lines is biased towards those galaxies with the highest star formation rates, as these galaxies
tend to have higher overall emission line fluxes.
There is evidence that there is a relationship between the star formation rate,
stellar mass, and gas-phase metallicity of galaxies, called the Fundamental 
Metallicity Relation \citep[FMR,][]{mannucci2010, andrews2013}. However, there has been significant discussion in the literature as to the strength, systematics, and even the very existence of this relation \citep{hughes2013,sanchez2013,telford2016, pistis2023}. Nevertheless, if we consider the effect that the FMR-as-observed would have on our sample, we may expect our local mass--metallicity relation 
to be biased low due to our need to 
detect the weak auroral lines to estimate a metallicity and the 
stated anti-correlation between gas-phase metallicity and star formation rate. \rrr{However, we note that as shown in \autoref{s:appendix:strongline}, we reach the same estimate of the mass-loading factor for most choices of strong line calibrators}.

By model design, incurring a systematic offset in gas-phase 
metallicity at fixed stellar mass in our reference sample will be 
partially accounted for by an offset in the same direction across the full 
sample. Furthermore, and quite apart from the model construction, our observed mass--metallicity relation is in good agreement with
literature relations at both $z\sim 0$ \citep{berg2012, andrews2013, cook2014} and with the measured rate at which the mass--metallicity relation evolves \citep{mactii}. Though these studies may also be subject to the same potential SFR--metallicity biases, the heterogeneity of depth of the datasets --- which directly controls the lowest \halpha{} luminosity, and thus SFR, that is included at a given distance --- makes it difficult for such a concordance to be reached if the impact of a fundamental metallicity relation strongly biased 
the resulting observed MZR. We conclude that the discrepancy between our measured mass-loading factor and those required by cosmological simulations is unlikely to be a result of a systematic bias in our method.

{}

\section{Conclusions}
In this work, we have used the background (non-SAGA host satellite) spectra 
collected by the SAGA survey to probe the star-forming cycle of low-mass 
galaxies in the $z<0.21$ Universe. 
The SAGA survey's limiting magnitude is around a magnitude deeper than 
past wide-field efforts ($m_{\rm lim,SAGA}=20.75$), 
making the \sagabg{} galaxies a novel window into the low-mass and low-redshift Universe. 

In this work, we examined the apparent evolution of the star-forming main sequence and the stellar mass-gas phase metallicity
relation within the last \tlbmax{} as a method to constrain the star formation cycle at low redshift. 
We constructed a simple ``\rrr{retrogression}'' model to trace the evolution of our sample as a function of redshift. In essence, rather than attempt to instantiate the galaxy sample from 
theoretical principles, we ask whether we can accurately predict the appearance of the sample in $M_\star-\text{SFR}-Z_{\text{O}}$ space as a function of redshift by evolving our lowest redshift sample backward in time. At each timestep, we: remove the mass in stars that were made during the redshift bin (equivalently $\Delta t$) given the galaxy's SFR, remove the yield in oxygen that was generated by those stars, and restore to the ISM both the mass that had been locked in those stars and the gas that had been expelled by star formation feedback as quantified by some value of the mass-loading factor, $\etam{}\equiv \dot M_{\rm out}/\text{SFR}$. 

In the main technical thrust of this work, we showed that:
\begin{itemize}
    \item Our model, which includes both the physical evolution of the \sagabg{} sample and the sample observability as imposed by the SAGA selection and detection criteria, can successfully predict the apparent redshift evolution of the sample in stellar $M_\star$-$Z_O$-SFR space (see \autoref{s:results:sfms} and \autoref{s:results:mzr}).
    \item Our best-fit model favors a positive evolution of the SFMS normalization (which will be presented in greater depth in \citetalias{otherpaper}) and a low, near-unity mass-loading factor at \etamest{} (\autoref{s:results:parameters}). 
\end{itemize}

The uncertainty on our estimate of the mass-loading factor is 
dominated by the relatively small number of galaxies for which we 
can measure a ``direct'' auroral line gas-phase metallicity. 
A large increase in the number of low-mass galaxies at appreciable redshift (i.e. $z\gtrsim0.03$) for which these weak auroral lines are confidently detected would allow a stellar mass dependence to 
be introduced into this model.

In the context of the mass-loading literature, we found that:
\begin{itemize} 
    \item  Our estimate of $\etam{}$ is consistent with recent observational results from studies that directly estimate the mass outflow rate from detections of extra-planar gas (\autoref{s:discussion:direct}).
    \item Our measure of $\etam{}$ is in good agreement with some, 
    but not all, estimates of the mass-loading factor made by 
    modeling the mass-metallicity relation or other aspects of
    galaxy chemical evolution (\autoref{s:discussion:indirect})
    \item   We find good agreement with $\etam{}$ measurements of single galaxy simulations (\autoref{s:discussion:simulations:ism}). However, our estimate of $\etam{}$ is significantly lower than cosmological zooms and big box simulations, which predict mass-loading factors of up to $\etam\approx 50$ over our stellar mass range (\autoref{s:discussion:simulations:cosmo}). 
\end{itemize}

The fact that we arrive at consistent values of the mass-loading factors compared to direct measurements despite approaching the $\etam{}$ measurement problem with a very different set of assumptions --- our \rrr{retrogression} model makes strong assumptions about short-term galaxy evolution and ISM mixing, while direct $\etam{}$ studies make strong assumptions about the geometry and kinematic structure of the outflows --- is a crucial step forward for consensus-building in a volume of parameter space where theoretical predictions differ by more than an order of 
magnitude. 

Our results join a growing body of both observational and theoretical work that indicate that low-mass galaxies are not as efficient at moving gas to large 
radii as has been claimed is required in some simulations to produce realistic galaxies in this mass range. Due to the importance of star formation and star formation feedback in low-mass galaxies, changing the efficiency at which outflows are launched would not only have implications for the chemical enrichment of dwarfs, but also for their stellar structure \citep{elbadry2016, kadofong2020c, kadofong2022a}, gas content \citep{elbadry2018,mancerapina2019,pandya2020}, and structural diversity \citep{mancerapina2020, kadofong2021, cardonabarrero2023}.

This paper represents a first effort using the SAGA background spectra, but the range of low-mass galaxy science that can be done is far from exhausted in this work. In \citetalias{otherpaper} of this series, we will extend the model 
presented here to consider the physical evolution of the low-mass star-forming main sequence and the 
implications of that evolution for constraints of galaxy evolution models at low redshift. The line measurements and 
abundance inferences will also be released as part of a future SAGA data release \citeprep{Mao} such that 
the information presented in this work will be accessible to the broader astronomical community. 

\acknowledgements{}
The authors thank Ulrich Steinwandel, Drummond Fielding, Lachlan Lancaster, Christopher Carr, \srr{Chang-Goo Kim, and Eve Ostriker} for thoughtful discussions that improved the quality of this manuscript. \rrr{The authors additionally thank the anonymous referee for their thorough and useful report.}

This research made use of data from the SAGA Survey (sagasurvey.org). The SAGA Survey is a spectroscopic survey with data obtained from the Anglo-Australian Telescope, the MMT Observatory, and the Hale Telescope at Palomar Observatory. The SAGA Survey made use of public imaging data from the Sloan Digital Sky Survey (SDSS), the DESI Legacy Imaging Surveys, and the Dark Energy Survey, and also public redshift catalogs from SDSS, GAMA, WiggleZ, 2dF, OzDES, 6dF, 2dFLenS, and LCRS. The SAGA Survey was supported was supported by NSF collaborative grants AST-1517148 and AST-1517422 and by Heising–Simons Foundation grant 2019-1402. 

Observations reported here were obtained in part at the MMT Observatory, a joint facility of the University of Arizona and the Smithsonian Institution. Data were also acquired at the Anglo-Australian Telescope (AAT) under programs A/3000 and NOAO 0144/0267. We acknowledge the traditional owners of the land on which the AAT stands, the Gamilaraay people, and pay our respects to elders past and present.

GAMA is a joint European-Australasian project based around a spectroscopic campaign using the Anglo-Australian Telescope. The GAMA input catalogue is based on data taken from the Sloan Digital Sky Survey and the UKIRT Infrared Deep Sky Survey. Complementary imaging of the GAMA regions is being obtained by a number of independent survey programmes including GALEX MIS, VST KiDS, VISTA VIKING, WISE, Herschel-ATLAS, GMRT and ASKAP providing UV to radio coverage. GAMA is funded by the STFC (UK), the ARC (Australia), the AAO, and the participating institutions. The GAMA website is http://www.gama-survey.org/ . 

This project used public data from the Sloan Digital Sky Survey (SDSS). Funding for the Sloan Digital Sky Survey IV has been provided by the Alfred P. Sloan Foundation, the U.S. Department of Energy Office of Science, and the Participating Institutions. The SDSS-IV acknowledges support and resources from the Center for High-Performance Computing at the University of Utah. The SDSS website is www.sdss.org.

The SDSS-IV is managed by the Astrophysical Research Consortium for the Participating Institutions of the SDSS Collaboration, including the Brazilian Participation Group, the Carnegie Institution for Science, Carnegie Mellon University, the Chilean Participation Group, the French Participation Group, Harvard-Smithsonian Center for Astrophysics, Instituto de Astrofísica de Canarias, The Johns Hopkins University, the Kavli Institute for the Physics and Mathematics of the Universe (IPMU)/University of Tokyo, the Korean Participation Group, Lawrence Berkeley National Laboratory, Leibniz Institut für Astrophysik Potsdam (AIP), Max-Planck-Institut für Astronomie (MPIA Heidelberg), Max-Planck-Institut für Astrophysik (MPA Garching), Max-Planck-Institut für Extraterrestrische Physik (MPE), the National Astronomical Observatories of China, New Mexico State University, New York University, the University of Notre Dame, Observatário Nacional/MCTI, The Ohio State University, Pennsylvania State University, Shanghai Astronomical Observatory, the United Kingdom Participation Group, Universidad Nacional Autónoma de México, the University of Arizona, the University of Colorado Boulder, the University of Oxford, the University of Portsmouth, the University of Utah, the University of Virginia, the University of Washington, the University of Wisconsin, Vanderbilt University, and Yale University.

This project used public data from the Legacy Surveys. The Legacy Surveys consist of three individual and complementary projects: the Dark Energy Camera Legacy Survey (DECaLS; NOAO Proposal ID No. 2014B-0404; PIs: David Schlegel and Arjun Dey), the Beijing-Arizona Sky Survey (BASS; NOAO Proposal ID No. 2015A-0801; PIs: Zhou Xu and Xiaohui Fan), and the Mayall z-band Legacy Survey (MzLS; NOAO Proposal ID No. 2016A-0453; PI: Arjun Dey). Together, DECaLS, BASS, and MzLS include data obtained, respectively, at the Blanco telescope, Cerro Tololo Inter-American Observatory, National Optical Astronomy Observatory (NOAO); the Bok telescope, Steward Observatory, University of Arizona; and the Mayall telescope, Kitt Peak National Observatory, NOAO. The Legacy Surveys project is honored to be permitted to conduct astronomical research on Iolkam Du'ag (Kitt Peak), a mountain with particular significance to the Tohono O'odham Nation.

The NOAO is operated by the Association of Universities for Research in Astronomy (AURA) under a cooperative agreement with the National Science Foundation.

The BASS is a key project of the Telescope Access Program (TAP), which has been funded by the National Astronomical Observatories of China, the Chinese Academy of Sciences (the Strategic Priority Research Program "The Emergence of Cosmological Structures" grant No. XDB09000000), and the Special Fund for Astronomy from the Ministry of Finance. The BASS is also supported by the External Cooperation Program of the Chinese Academy of Sciences (grant No. 114A11KYSB20160057) and the Chinese National Natural Science Foundation (grant No. 11433005).

The Legacy Survey team makes use of data products from the Near-Earth Object Wide-field Infrared Survey Explorer (NEOWISE), which is a project of the Jet Propulsion Laboratory/California Institute of Technology. NEOWISE is funded by the National Aeronautics and Space Administration.

The Legacy Surveys imaging of the DESI footprint is supported by the Director, Office of Science, Office of High Energy Physics of the U.S. Department of Energy under contract No. DE-AC02-05CH1123; the National Energy Research Scientific Computing Center, a DOE Office of Science User Facility under the same contract; and the U.S. National Science Foundation, Division of Astronomical Sciences under contract No. AST-0950945 to NOAO.

This project used public archival data from the Dark Energy Survey (DES). Funding for the DES Projects has been provided by the U.S. Department of Energy, the U.S. National Science Foundation, the Ministry of Science and Education of Spain, the Science and Technology Facilities Council of the United Kingdom, the Higher Education Funding Council for England, the National Center for Supercomputing Applications at the University of Illinois at Urbana-Champaign, the Kavli Institute of Cosmological Physics at the University of Chicago, the Center for Cosmology and Astro-Particle Physics at the Ohio State University, the Mitchell Institute for Fundamental Physics and Astronomy at Texas A\&M University, Financiadora de Estudos e Projetos, Fundação Carlos Chagas Filho de Amparo à Pesquisa do Estado do Rio de Janeiro, Conselho Nacional de Desenvolvimento Científico e Tecnológico and the Ministério da Ciência, Tecnologia e Inovação, the Deutsche Forschungsgemeinschaft, and the Collaborating Institutions in the Dark Energy Survey.

The Collaborating Institutions in the Dark Energy Survey are Argonne National Laboratory, the University of California at Santa Cruz, the University of Cambridge, Centro de Investigaciones Energéticas, Medioambientales y Tecnológicas-Madrid, the University of Chicago, University College London, the DES-Brazil Consortium, the University of Edinburgh, the Eidgenössische Technische Hochschule (ETH) Zürich, Fermi National Accelerator Laboratory, the University of Illinois at Urbana-Champaign, the Institut de Ciències de l'Espai (IEEC/CSIC), the Institut de Física d'Altes Energies, Lawrence Berkeley National Laboratory, Ludwig-Maximilians Universität München and the associated Excellence Cluster Universe, the University of Michigan, the National Optical Astronomy Observatory, the University of Nottingham, The Ohio State University, the OzDES Membership Consortium, the University of Pennsylvania, the University of Portsmouth, SLAC National Accelerator Laboratory, Stanford University, the University of Sussex, and Texas A\&M University.

The public archival data from the DES are based in part on observations at Cerro Tololo Inter-American Observatory, National Optical Astronomy Observatory, which is operated by the Association of Universities for Research in Astronomy (AURA) under a cooperative agreement with the National Science Foundation.

\software{Astropy \citep{astropy:2013, astropy:2018}, matplotlib \citep{Hunter:2007}, SciPy \citep{jones_scipy_2001}, the IPython package \citep{PER-GRA:2007}, NumPy \citep{van2011numpy}, 
pandas \citep{McKinney_2010, McKinney_2011},
Astroquery \citep{astroquery}, extinction \citep{barbary2021}, WebPlotDigitizer \citep{rohatgi2022}}

\bibliography{sagabg.bib,sagabg_paperi.bib,software.bib,gama.bib}

\appendix{}
\iffirst{
\section{Flux Calibration and Line Measurement of the Spectra}\label{s:appendix:fluxcalibration}
To ensure that we achieve reasonable line fits, we test our flux recovery for a sample of galaxy spectra with reliable spectroscopic redshifts (\textsf{nQ}$>2$ in the GAMA DR4 \textsf{StellarMassesLambdarv20} catalog) from the GAMA survey at $0<z<0.5$ with $M_r>-10$. We only consider the flux-calibrated spectra in the DR4 release (i.e. those originating either from GAMA or SDSS). Both our measurements and the GAMA DR4 catalog (\textsf{GaussFitComplexv05}) should be corrected for stellar absorption, but not for reddening. 

In \autoref{f:gamafluxes} we show a comparison between the GAMA catalog \halpha{} flux measurements and our analogous flux measurements. We find an overall excellent agreement, though there is a tail wherein our measured flux is lower than the GAMA catalog values (i.e. below the 1:1 line in \autoref{f:gamafluxes}). This may be due to differences in how \halpha{} absorption is handled, as \halpha{} absorption is not typically resolved in \aat{} spectra for these galaxies.

\begin{figure}[htb]
  \centering     
  \includegraphics[width=.5\linewidth]{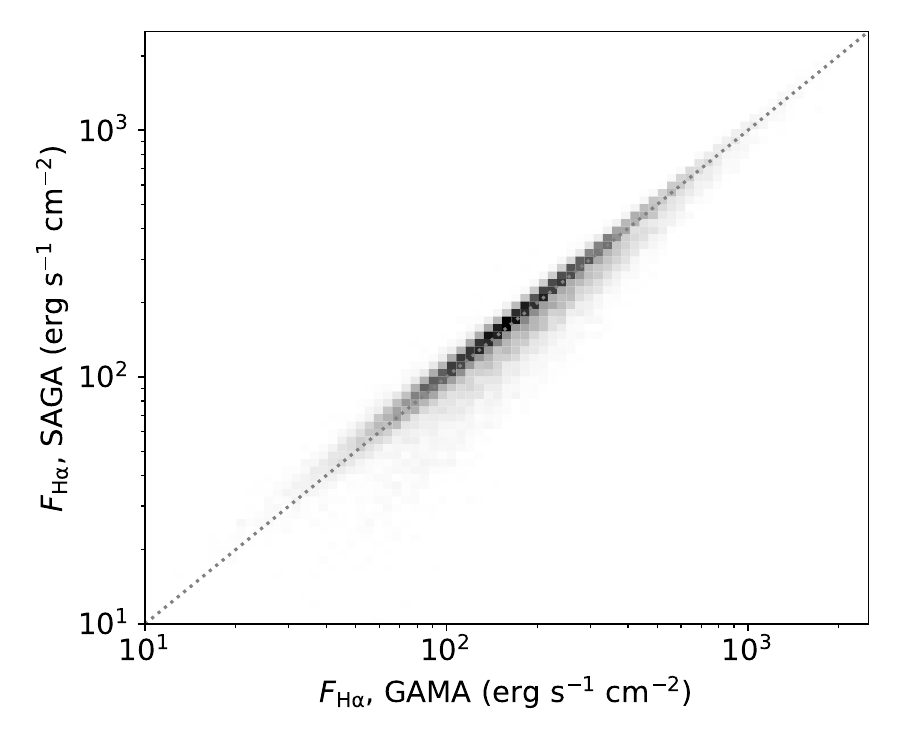}
  \caption{ 
      A comparison between our measured \halpha{} fluxes and GAMA DR4 catalog fluxes for 
      GAMA galaxies in our stellar mass and redshift range. Both measures are nominally corrected for
      stellar absorption, but not for dust. We show here only galaxies with an \halpha{} SNR greater than 3 with a positive \halpha{} flux error in the GAMA DR4 catalog.
      }\label{f:gamafluxes}
\end{figure}

We additionally test that our spectra flux calibration is accurate by comparing to \sagabg{} targets for which both SAGA \aat{}/\mmt{} and GAMA DR4 spectra were obtained. In \autoref{f:fluxcalline} we compare 
absorption-corrected \halpha{} fluxes from this overlap, discarding galaxies where $A_\lambda(H\alpha)>3$ or $L_{H\alpha} < L_{H\alpha,\lim}$ in our SAGA spectra measurements. We find a good agreement between the two surveys across more than an order of magnitude in flux space.

\begin{figure}[htb]
  \centering
  \includegraphics[width=.5\linewidth]{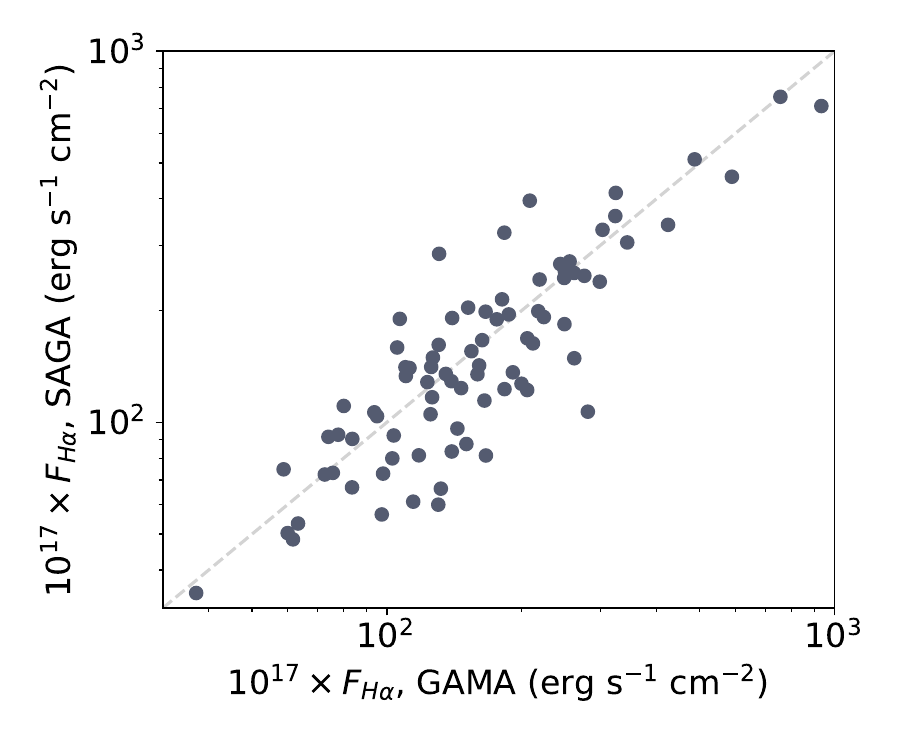}
  \caption{ 
      \halpha{} fluxes as measured from GAMA DR4 versus \halpha{} fluxes for the same galaxies as measured from SAGA \aat{} spectra. 
      Note that these are spectra for which the underlying source spectrum is \textit{not} the same, as opposed to \autoref{f:gamafluxes}.
      }\label{f:fluxcalline}
\end{figure}

{} 
}
\begin{figure*}[b]
  \centering     
  \includegraphics[width=\linewidth]{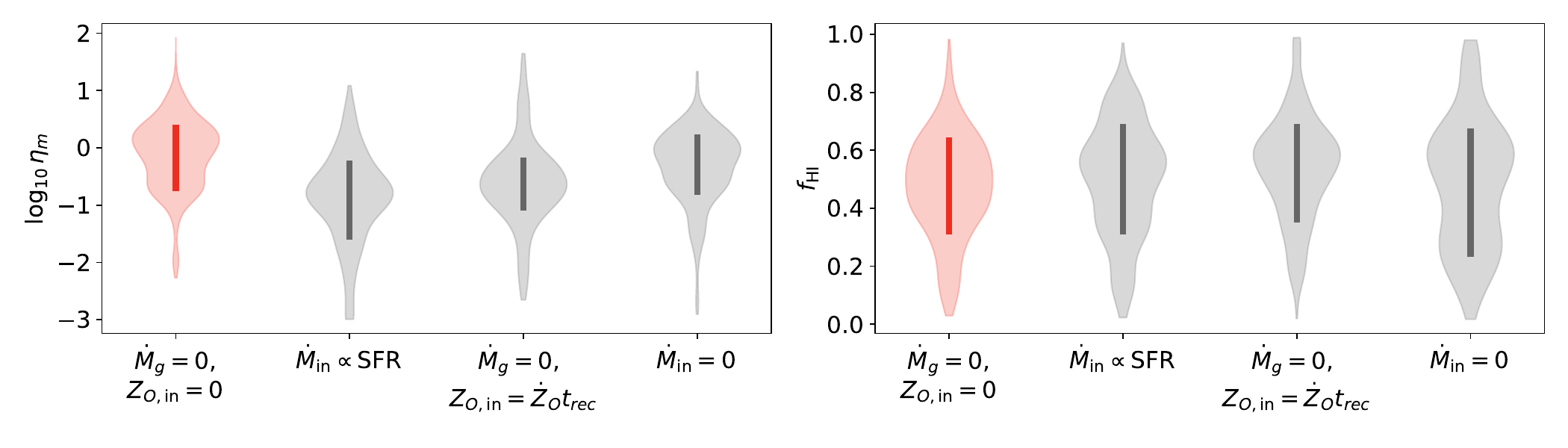}
  \caption{ 
      A comparison in inferred parameter distributions for our fiducial model (red, $\dot M_{g}=0$) and three alternate models for gas accretion (grey).
      \rrr{The vertical line in each distribution shows the range from the 16\tth{} to the 84\tth{} percentile of the posterior.}
      We consider a model where accretion is proportional to SFR (middle distribution in each panel), where accretion balances mass loss in the gas reservoir using gas from the CGM (second from right in each panel), and where there is no accretion of gas at all (rightmost in each panel). In all cases, we find no evidence for a systematic
      shift in our inferred parameters beyond the fiducial inference uncertainties.
      }\label{f:accretioncomparison}
\end{figure*}

\section{Alternate Models of Gas Inflow}\label{s:appendix:withaccretion}
Gas accretion in the last \tlbmax{} onto the low-mass galaxies in our sample is an interesting if poorly constrained component of the baryon cycle. 
Though understanding these inflows is of significant scientific interest, in this work we will simply interrogate whether our assumptions about accretion will significantly affect the outcome of our inference. 
Another way to frame this section is to ask whether our inference 
can inform our understanding of recent gas inflows into low-mass galaxies; we conclude that the inference is not sensitive to choices about inflows, and leave the editorial evaluation of whether that is a boon or drawback to the reader.

As mentioned in the main text of \autoref{s:modeling}, we consider three models for gas accretion in this Appendix. First, we consider a model where 
the rate of pristine gas inflow scales with the SFR, i.e.
\begin{equation}
  M_{\rm in} = \alpha \dot M_{\star},
\end{equation}
where $\alpha$ is an additional free parameter in our inference. This 
relationship between inflow and SFR has been suggested in the literature 
for disks in equilibrium \citep{schmidt2016,krumholz2018}.

Next, we consider a model in which no inflow is included, i.e. $\dot M_{\rm in}=0$. This will induce a negative
time derivative of gas mass, $\dot M_g<0$, since gas is either converted to stellar mass or expelled from the galaxy.

For the last inflow models, we consider the steady-state solution, following \cite{fielding2017} in defining the steady-state set-up for gas accretion as in our fiducial model:
\begin{equation}
  \begin{split}
    \dot M_{\rm in} &= \dot M_{\rm out} + \dot M_{\star} \\ 
            &= (1 + \etam{}) \dot M_{\star},
  \end{split}
\end{equation}
which, given that $\dot M_g = \dot M_{\rm in }-\dot M_{\rm out} - \dot M_{\star}$ implies that $\dot M_g=0$ when we recall that $\etam{}  = \dot M_{\rm out}/\dot M_{\star}$. We propose two variations on this simple accretion model: one where the incoming gas is pristine ($Z_O=0$, the cosmological inflow case), and one where the incoming gas is cooled from the CGM. It is both theoretically expected
and observationally realized that the metallicity of the CGM should be 
generally lower than that of the ISM \citep{kacprzak2019}; we adopt a simple
approximation for the relationship between the metallicity of the ISM and
CGM wherein the metallicity difference is approximately equal to the 
expected change in oxygen abundance of the ISM over a characteristic recycling time. This recycling time is the timescale over which fountain flows launched into the CGM
from the ISM are expected to cool and accrete back onto the ISM; we 
adopt $t_{rec}=0.5$ Gyr following the results of \cite{anglesalcazar2017} 
in the relevant mass range. 

In all three cases, we find that our inference of $\etam{}$ and $\fhi{}$ does not shift beyond the 16\tth{} to 84\tth{} percentile region of our 
fiducial posterior distributions, as shown in \autoref{f:accretioncomparison}. Our estimate of $\asfh{}$ is also unaffected; this parameter is not expected to be strongly affected by changes in gas inflow prescriptions. We thus conclude that the results presented in this work 
are not sensitive to and cannot distinguish between models of recent 
gas inflow. Part of this may be that the recycling time of fountain flows
launched from these galaxies into the CGM is not long compared to the
timesteps (or redshift bins) adopted in the study. Thus, much of the gas 
that will be reaccreted into the ISM from the CGM has already done so in the $\Delta t_{lb}\sim 400$ Myr that defines our movement from one redshift bin to the next. 

\begin{figure*}[b]
  \centering     
  \includegraphics[width=\linewidth]{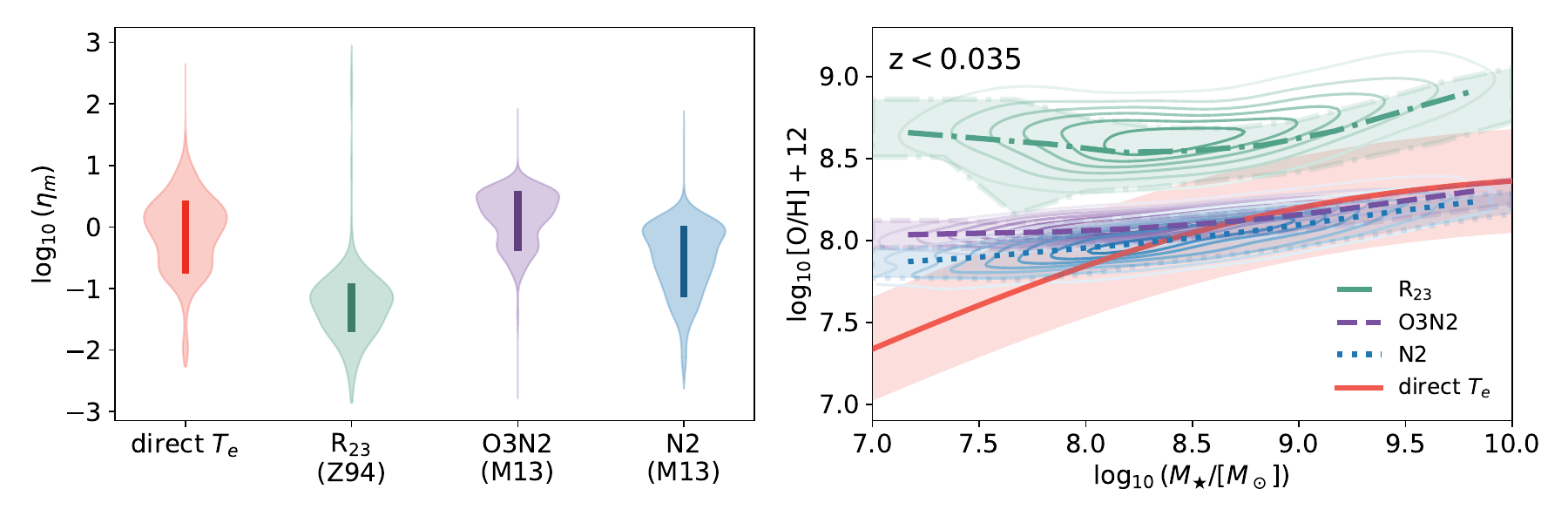}
  \caption{ 
      \rrr{
      \textit{Left:} A comparison of the effect of our choice of metallicity calibrator on our inferred mass-loading factor (left) and 21cm-bright gas fraction (right). As in \autoref{f:accretioncomparison}, we show our fiducial results in red and our alternate metallicity calibrators in grey. We show the ${\text R_{23}}$ calibration of \cite{zaritsky1994} at center and the O3N2 calibration of \cite{marino2013} at right. We find that the O3N2 strongline calibration produces a mass-loading factor that is consistent with our direct-\te{} fiducial model ($\etam(\text{O3N2}) = 1.8^{+2.1}_{-1.3}$), and that the ${\text R_{23}}$ calibration produces a mass-loading factor somewhat lower than our fiducial model ($\etam(\rm R_{23}) = 0.06^{+0.06}_{-0.04}$). \textit{Right:} The $z<0.035$ mass-metallicity relation for each calibrator as compared to our mass-metallicity fit from \autoref{f:localMZR}, which has been validated against auroral line metallicity estimates from the local Universe.
      }
      }\label{f:rrr_stronglines}
\end{figure*}

\section{\rrr{Alternate Gas-Phase Metallicity Estimates}}\label{s:appendix:strongline}
\rrr{
    In addition to alternative models of inflow, it is useful to understand the impact of our choice of metallicity calibration on our estimate of the mass-loading factor. In our fiducial model, we use direct temperature metallicity estimates in order to alleviate potential biases resulting from the considerable offsets present between different strongline calibrations in the literature (for an overview, see \citealt{kewley2019}). However, as discussed in \autoref{s:measurements}, adopting a metallicity estimation method that relies upon the detection of the weak auroral lines may introduce its own biases.
}

\rrr{To understand the implication of using direct temperature metallicities on our estimate of the mass-loading factor, we rerun our fiducial model using three different strong-line ratio calibrations from the literature. We use three strong-line ratios for this exercise: $\rm R_{23} \equiv [\lineflux{[OII]\lambda3727\rm \AA} + \lineflux{[OII]\lambda3729\rm \AA} + \lineflux{[OIII]\lambda4959\rm \AA} + \lineflux{[OIII]\lambda5007\rm \AA}]/\lineflux{H\beta}$, O3N2 $\equiv [\lineflux{[OIII]\lambda5007\rm \AA}/\lineflux{H\beta}]/[\lineflux{[NII]\lambda 6583\AA}/\lineflux{H\alpha}]$, and N2 $\equiv \lineflux{[NII]\lambda 6583\AA}/\lineflux{H\alpha}$. We adopt the $\rm R_{23}$ calibration of \cite{zaritsky1994} and the O3N2 and N2 calibrations from \cite{marino2013}:}
    \rrr{
    \begin{subequations}
        \begin{equation}
            (\log_{10}[\text{O/H}] + 12)_{\text{Z94}} = 9.265 - 0.33\log_{10}(\rm R_{23}) - 0.202 \log_{10}(\rm R_{23})^2 - 0.207 \log_{10}(\rm R_{23})^3 - 0.333 \log_{10}(\rm R_{23})^4
        \end{equation}
        \begin{equation}
            (\log_{10}[\text{O/H}] + 12)_{\rm M13,N2} = 8.743 + 0.462 \log_{10}(\text{N2}) 
        \end{equation}
        \begin{equation}
            (\log_{10}[\text{O/H}] + 12)_{\rm M13,O3N2} = 8.533 - 0.214 \log_{10}(\text{O3N2})
        \end{equation}
    \end{subequations}
    }
    \rrr{We propagate errors on the metallicities via standard propagation of errors from the uncertainty on the measured line fluxes. Only metallicities derived from galaxies whose \hbeta{} and \ionline{O}{3}{4959} fluxes are measured with a $\text{SNR}>5$ are included in the model; the remaining galaxies at $z<0.035$ are assigned metallicities from the best-fit mass-metallicity relation as in the fiducial model.}

    \rrr{In \autoref{f:rrr_stronglines} we show the posterior distributions of $\etam{}$ and $\fhi{}$ for both our fiducial run (red and left in each panel) and the strong-line calibration runs (grey). We find that the two strong-line metallicity calibrations of \cite{marino2013} produce estimates of $\etam{}$ that are in good agreement with our fiducial model ($\etam(\text{O3N2}) = 1.8^{+2.1}_{-1.3}$ and $\etam(\text{N2}) = 0.35^{+0.78}_{-0.35}$), while the $R_{23}$ calibration of \cite{zaritsky1994} is best-fit by a somewhat smaller mass-loading factor ($\etam(\rm R_{23}) = 0.06^{+0.06}_{-0.04}$). The $\rm R_{23}$-based estimate of the mass-loading factor is in mild tension with our fiducial result: $\langle \etam(\rm T_e) \rangle_{10} \approx \langle \etam(\rm R_{23}) \rangle_{87}$ where $\langle \rangle_{X}$ denotes the $X^{\rm th}$ percentile of the posterior.}

    \rrr{The tension between the $\rm R_{23}$-based calibration and the other methods can be understood when we consider that the \cite{zaritsky1994} calibration predicts gas-phase metallicities $\sim 0.5$ dex higher than the direct temperature metallicity method and \cite{marino2013} calibrations (see also Figure 2 of \citealt{kewley2008}), as shown in the right panel of \autoref{f:rrr_stronglines}. If we take the limiting case where the strong-line calibrations simply represent a constant shift in oxygen mass fraction, it is straightforward to show that a positive shift in $Z_O$ (that is, $Z_O({\rm Z94}) = Z_O({\rm T_e}) + \Delta Z_O({\rm Z94})$ where $\Delta Z_O({\rm Z94})>0$) will result in a lower estimated mass-loading factor if all other parameters are held fixed.}

    \rrr{From this exercise we have demonstrated two points salient to the main analysis of this work. First, we have shown that strong-line calibrators produce mass-loading factors consistent with or lower than the direct temperature method. This is especially important when we consider that theoretical predictions of the mass-loading are, in general, significantly higher than observational estimates. The mass-loading factor produced by the strong-line calibrator that produces the results that diverge the most from our fiducial $\etam$ estimate does not favor the high mass-loading factors used in many contemporary simulations of galaxy evolution (see \autoref{s:discussion:simulations}). Second, we find that the absolute calibration of our metallicity estimates does significantly impact our inferred mass-loading factor. This finding supports our decision to use the direct temperature metallicity method to estimate gas-phase metallicities in the main text of this work, as direct-\telec{} metallicities can be more easily validated against analogous results from the literature and do not suffer from the large offsets seen in strong-line calibrations.}

\section{Inferring Total Gas Masses via a fixed Star Formation Efficiency}\label{s:appendix:gasmass}

A galaxy's gas reservoir and its star formation are, of course, causally linked. In our fiducial approach, we estimate \HI{} mass via the measured $M_\star-M_{\rm HI}$ relation of 
\cite{bradford2015}. Another way that one may consider assigning \HI{} masses to the galaxies in our reference sample is by assuming the star formation efficiency (here defined as $\epsilon = \text{SFR}/M_{\rm HI}$) of the galaxies rather than their $M_\star - M_{\rm HI}$ relation. Star formation efficiency is relatively constant across the stellar mass range probed by our sample, albeit with significant scatter \citep{huang2012,durbala2020}

Neither method of assigning $M_{\rm HI}$ to galaxies is perfect: assuming the \cite{bradford2015} relationship between stellar mass and \HI{} mass overestimates the variance in the \HI{} mass distribution at fixed stellar mass and SFR due to unaccounted-for covariance between \HI{} mass and SFR, while assuming a fixed star formation efficiency places a 
very strong assumption about the relationship between past and current star formation by directly linking the total fuel for star formation to the star formation at $z=0$.
Moreover, it has been shown that the overall structure of low-mass galaxies contributes
significantly to their star formation efficiency \citep[see, e.g.][]{delosreyes2019, ostriker2022}; assuming the same SFE for all galaxies in our reference sample would also be an oversimplification.
As an exercise in understanding the implications of our choice, we have rerun our inference using a 
fixed SFE where $\log_{10}\epsilon \sim \mathcal{N}(-10,0.5)$ based on SFE measurements 
by~\cite{huang2012} and find that it does not affect 
the outcome of this analysis.

\section{\rrr{Observed and Predicted Summary Statistics in $M_\star$-$Z_O$-SFR-\lowercase{z} space}}\label{s:appendix:validation}

To supplement the model optimization described in \autoref{s:modeling} and shown in \autoref{f:SFMS_evolution} 
and \autoref{f:MZR_evolution}, we consider a brief exercise in model validation below. In \autoref{f:samplecomparison}, we simply 
show the \textit{observed} redshift evolution of the median of each of our observable quantities --- 
stellar mass, star formation rate, and gas-phase metallicity --- for our observed sample and model predictions.

Here we are considering the total observed evolution, and the trends shown in \autoref{f:samplecomparison} are 
not indicative of the underlying physical shift in the median value of the measured quantities. We find that our model (blue) is able to explain the trend in the observed redshift evolution of the median star formation rate (middle) and gas-phase metallicity (right) to the precision of the 90\% confidence interval on the sample median. The observed evolution of the stellar mass does deviate from the predicted sample evolution by more than the extent of the 90\% confidence interval, but because the offset is small relative to the 
stellar mass range probed we conclude that the model is sufficient to describe the observed evolution of 
our sample in $M_\star$-SFR-$Z_O$ space.

\begin{figure*}[t]
  \centering     
  \includegraphics[width=\linewidth]{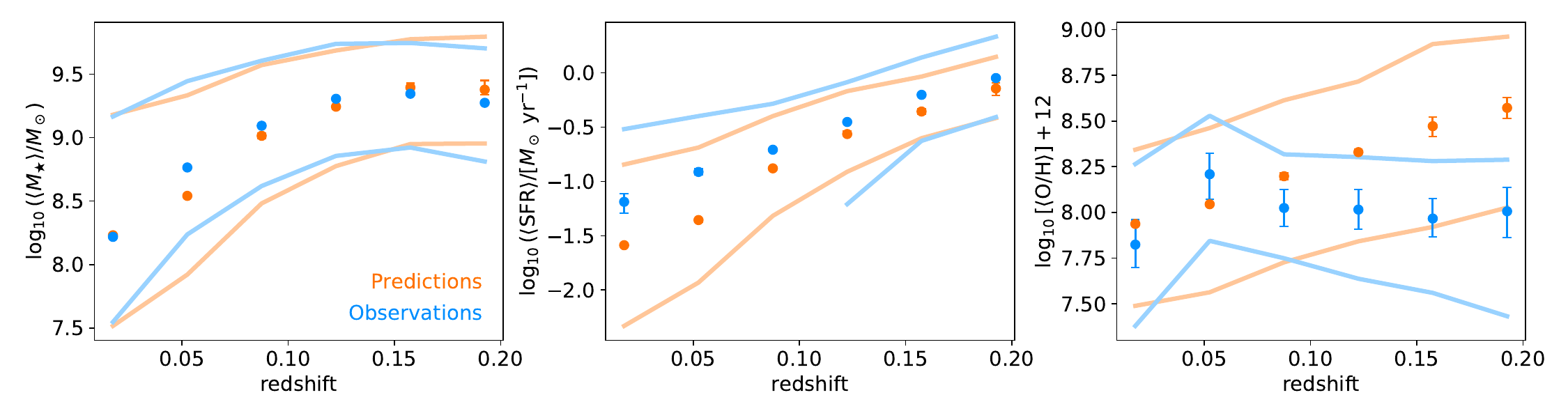}
  \caption{ 
      Redshift versus the observed (blue) and predicted (orange) samples in stellar mass (left), 
      star formation rate (middle), and gas-phase metallicity (right). In each panel, the 
      errorbars show the 90\% confidence interval on our estimate of the median and the curves indicate the 
      16\textsuperscript{th} and 84\textsuperscript{th} percentiles of the distribution. We find that our 
      model successfully traces the observed shift in our observed sample as a function of redshift, though 
      we stress to the reader that this figure reflects observational rather than physical redshift evolution.
      }\label{f:samplecomparison}
\end{figure*}

\section{Aperture Effects}
One observational uncertainty that has persisted across studies of metallicity relations at higher masses has been the fiber covering fraction of the targets. 
The problem of fiber coverage is ameliorated in the case of the SAGA galaxies due to the small on-sky sizes of our targets (compared to, for example, the higher mass SDSS sample of \citealt{mannucci2010}). 
It is nevertheless important to understand to what degree an aperture effect could influence our results. To do so we perform a simple partial correlation analysis --- that is, we 
consider whether there is evidence for a correlation between gas-phase metallicity and redshift
beyond that which can be explained by correlations between each variable and on-sky size. 

We expect to see a correlation between both on-sky size and metallicity (due to the correlation between size and stellar mass) and between on-sky size and redshift (due to the relation between angular diameter distance and redshift). 
To test whether redshift and gas-phase metallicity are correlated beyond what would be implied by their correlations with on-sky size, 
we compute the linear regression of gas-phase metallicity with respect to 
on-sky size (left panel of \autoref{f:aperture_correlation}) and the linear regression 
of redshift with respect to on-sky size (middle panel of \autoref{f:aperture_correlation}). 
If the observed correlation between redshift and gas-phase metallicity is driven only 
by fiber coverage fraction, then we would expect the residuals of these linear regressions to be uncorrelated.
Here we restrict our sample to galaxies with stellar masses exceeding \logmstar[$>8.9$], the median 
stellar mass of our sample with auroral line metallicity measurements. We do so to probe a 
narrower range of the mass--metallicity relation, which we expect to lessen the correlation 
between all three variables considered due to lower mass galaxies dropping out of the sample. 

We see a moderate negative correlation (r=-0.45, in line with the trend seen in the main body of this work), indicating that redshift and gas-phase metallicity are
correlated independently of on-sky size. We also computed the linear regression 
of gas-phase metallicity and on-sky size with respect to redshift to test whether there is 
evidence for on-sky size and gas-phase metallicity being correlated independently of redshift
and found that the correlation is very weak (r=-0.08). From this test, we conclude that our 
results are not due to a fiber coverage fraction that varies with redshift.

\begin{figure*}[htb]
  \centering     
  \includegraphics[width=\linewidth]{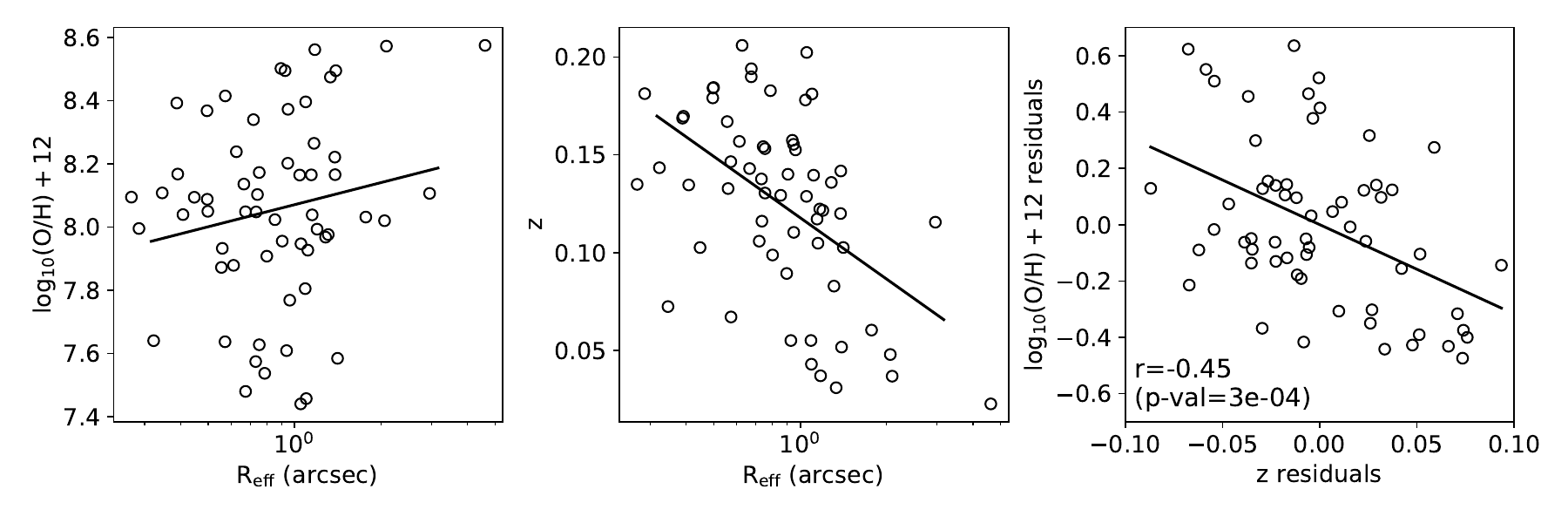}
  \caption{ 
      The results of a partial correlation analysis to demonstrate that redshift and 
      gas-phase metallicity are correlated independently of each variable's correlation with
      on-sky size. At left we show effective on-sky radius versus gas-phase metallicity,  
      where the black line shows the result of a linear regression of gas-phase metallicity 
      with respect to effective radius. In the middle panel, we show the same but for 
      redshift and effective radius. At right, we show the residuals of each linear regression;
      if the correlation between redshift and gas-phase metallicity come from a correlation 
      with effective radius then these residuals should be uncorrelated. We find evidence 
      for a moderate negative correlation (r=-0.45, p-value$<10^{-3}$ where the null 
      hypothesis is that the residuals are uncorrelated and normally distributed).       
      In each panel, we show only galaxies with stellar masses exceeding \logmstar[$>8.9$]. 
      }\label{f:aperture_correlation}
\end{figure*}

\iffirst{
\section{The Contribution of Environment}\label{s:appendix:environment}
The main scope of this paper is not to assess the detailed environment of the SAGA background galaxies, but it is important to our analysis to
assess whether the environment of the galaxies in our sample is changing as a function of redshift. The overall satellite fraction for galaxies in a given stellar mass range is not
expected to change significantly over $0<z<0.21$ \citep{conroy2006, mandelbaum2006, guo2019,  shuntov2022}, but we would like to confirm that this is true for the galaxies 
in our photometric selection.

To roughly assess the contribution of satellites to our sample, we consider only the SAGA background spectra that overlap the GAMA G09, G12, or G15 
regions. We then look for potential hosts of the SAGA galaxies from both the GAMA and NASA Sloan Atlas (NSA) catalogs, where we define a potential 
host as a galaxy with $M_r<-22$ using the source catalog photometry that lies within a projected distance of 2.5 Mpc of the SAGA galaxy with
a velocity difference of $\Delta v<1000\ \kms$.  
We expect GAMA to be complete over our redshift range given the cut in absolute magnitude 
\citep{baldry2018}, but the overlapping SAGA hosts are close to or cross the edge of the GAMA footprint. We thus expect the 
GAMA hosts to be complete only at around $z>0.14$, where the projected comoving distance per 0.5 deg is twice our projected distance cut. 
Conversely, although the NSA is complete for our low-redshift hosts, the catalog only extends out to $z=0.055$. 

We can nevertheless use this combination of datasets to determine whether the satellite fraction (the fraction of galaxies that are a 
satellite) changes significantly with redshift. In \autoref{f:satfraction} we show the satellite fraction as a function of redshift, where the
region at $0.055<z<0.14$ is covered by a grey shaded hatch to indicate the region of GAMA and NSA host catalog incompleteness. We find that the
satellite fraction in our sample is consistent with a constant value of around 6\%. 
This is somewhat lower than the 
satellite fractions found for the full low-mass galaxy population in simulations \citep{kravtsov2004, conroy2006, mandelbaum2006, guo2019,  shuntov2022}, which is 
consistent with the well-established idea that the quenched low-mass galaxy population is dominated by satellites \citep{geha2012}. 

For the purview of this analysis, the salient point of the exercise above is that the satellite fraction is consistent with a low and
constant value over the redshift range considered in this work. Thus, we expect that neither differential evolution of star-forming satellite galaxies nor the 
overdensity of satellites associated with the SAGA hosts at $z<0.01$ have influenced the results presented in this work. 

\begin{figure*}[t]
  \centering     
  \includegraphics[width=0.4\linewidth]{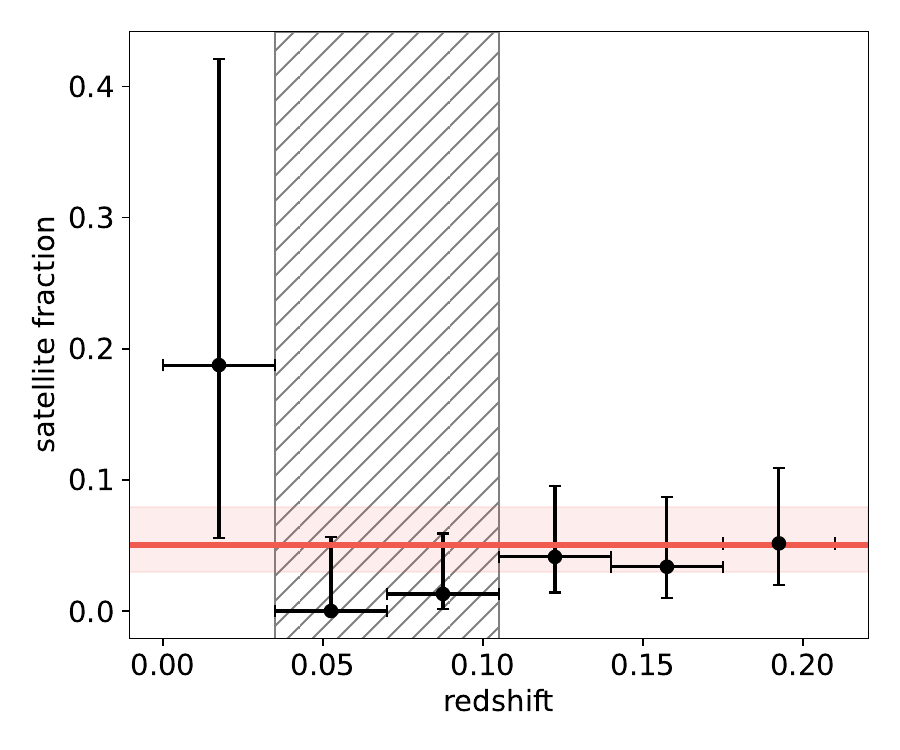}
  \caption{ 
      The estimated fraction of SAGA background galaxies in the present work that are satellites as estimated from a joint 
      match to the GAMA and NSA catalogs for galaxies at $M_r<-22$. We shade the region at $0.055<z<0.14$ to show the redshift space
      where neither NSA nor GAMA is complete. The uncertainties on the satellite fraction are computed as confidence intervals for a 
      binomial population using a Jeffreys Bayesian interval. The red line and shaded region show the satellite fraction computed across 
      $(z<0.055)\lor(z>0.14)$.
      }\label{f:satfraction}
\end{figure*}
}

\end{document}